\documentstyle[aps,preprint,epsf]      {revtex}
\begin{document}
\draft
\setcounter{page}{1}
\pagestyle{plain}

\title{\bf\Large Singularities in the Fermi liquid
description of a partially filled Landau level and
 the energy gaps of  fractional quantum Hall states}
\author{Ady Stern\cite{byline} and Bertrand I. Halperin\\
Physics Department, Harvard University, Cambridge, Massachusetts 02138}

\date{\today}
\maketitle
\noindent
\bigskip

\begin{abstract}
We consider a two dimensional electron system in an external magnetic field
at and near an even denominator
Landau level  filling fraction. Using a fermionic
Chern--Simons approach we study the description of the system's low
energy excitations within an extension of Landau's Fermi liquid theory.
We calculate perturbatively
 the effective mass and the quasi--particle interaction function
 characterizing this
description.  We find that at an
even denominator filling fraction the
fermion's effective mass diverges logarithmically
at the Fermi level, and argue that
this divergence allows for an {\it exact} calculation of the energy gaps
of the fractional quantized Hall states asymptotically approaching these
filling
fractions. We find that the quasi--particle interaction
function approaches a delta function.
 This singular behavior
  leads to a cancelation of the diverging effective mass from the
long wavelength low frequency linear response functions at even
denominator filling fractions.

\end{abstract}
\vfill\eject

%\begin{multicols}{1}
\narrowtext
\section{Introduction}

The Fermionic Chern--Simons theory has been extensively used in the last
few years to describe
the physics of a two dimensional electron gas in a partially filled
Landau level \cite{Hlr}
-- \cite{Aim}.
In that theory, the problem of electrostatically
interacting electrons in a strong magnetic field is converted, using an
exact transformation, to a problem of electrostatically
interacting "transformed fermions"  which also interact
with a Chern--Simons gauge field.

Being still unsolvable, the transformed fermion problem is studied
by approximations. The simplest approximation, mean field theory, already
gives interesting physical results. For example, it predicts the
existence of stable quantized Hall states at filling fractions of the
form $\frac{p}{\tilde{\phi}p+1}$, where $p$ is an arbitrary integer,
and $\tilde\phi$ is an even positive integer, equal to the number of
quanta of Chern--Simons flux attached to each fermion. As was first noted
by Jain
\cite{Jain}, filling fractions of the above form, with $\tilde{\phi}=2$,
are the most prominent fractional Hall states seen in the lowest Landau
level. (Jain's "composite fermions" may be regarded as equivalent to the
transformed fermions, projected onto the lowest electronic Landau level.)
By contrast, at the even denominator fractions $1/\tilde{\phi}$,
the fermion Chern--Simons mean field theory predicts a Fermi--liquid
state, with no energy gap and no quantized Hall effect \cite{Hlr}\cite{Kz}.

For a description of the excitation spectrum and of other quantities of
interest, it is of course necessary to go beyond mean field theory.
Specifically, it is necessary to consider, using perturbation theory, the
effects of fluctuations in the Chern--Simons field and of the two
body interaction, effects which are omitted from mean field theory.
Consequences of these perturbations were explored at some length by
Halperin, Lee and Read (HLR)\cite{Hlr} for the Fermi liquid state at
$\nu=1/\tilde{\phi}$, and also to some degree for other nearby filling
fractions.

An interesting result of the HLR analysis was the prediction that
at $\nu=1/\tilde{\phi}$ the
effective mass $m^*$ of the transformed fermions diverges for
quasi--particles at the Fermi energy. For the physically relevant case
of Coulomb interaction between the electrons, this divergence was
found to be quite weak, i.e., logarithmic in $\omega$, the energy relative
to the Fermi energy.
For short range electron--electron interaction HLR found a stronger
divergence $m^*\sim\omega^{-1/3}$, while for a longer range interaction
the effective mass remains finite as $\omega\rightarrow 0$.

The HLR results were based on an examination of the leading diagrams in the
perturbation expansion for small $\tilde{\phi}$, supplemented by heuristic
arguments that the leading singularity might not be affected by higher
orders in perturbation theory. Consequently, it is of interest to
see whether these arguments can be made more convincing, or whether by
contrast results contradictory to HLR may be obtained. Interest in these
questions has been stimulated by the fact that similar effective mass
divergences are found in other models in which fermions interact strongly
with a fluctuating gauge field, including models that have been advanced in
the context of high $T_c$ superconductivity \cite{Marston}--\cite{Misc}.

Assuming that the effective mass divergence predicted by HLR is indeed correct,
one may ask what physically measurable quantities might reflect it. According
to HLR, the renormalized mass should $not$ show up in the
long wavelength density and current
response functions at zero temperature, for $\omega /q\rightarrow  0$. This
observation was confirmed, more recently, by analyses using a variety of
theoretical techniques, including renormalization group methods
\cite{Wilczek1}\cite{Wilczek2},
bosonization \cite{Marston}
and diagrammatic analyses\cite{Kim}. Moreover, Kim {\it et.al.}\cite{Kim},
using
an approximation that takes into account the leading divergent contribution to
$m^*$, have shown that there is no corresponding singularity in the density
response function in the limit of low frequencies and long wavelengths, for
{\it any} value of $\omega/q$. Within their approximation, the density response
function looks just like the one obtained in the random phase approximation
of a Fermi gas with the bare mass $m$. The diverging effective mass may be
expected to show up in the temperature dependence of the specific heat at
low temperatures; however, a contribution from the low frequency density
relaxation collective mode could lead to the same anomalous temperature
dependence as the quasiparticle contribution.

As noted by HLR, a more definitive
manifestation of the diverging $m^*$ should come from the behavior of the
energy gap at quantized Hall steps for $\nu =\frac{p}{\tilde{\phi}p+1}$, for
large values of $p$. Specifically, one might expect the energy gap $E_g(\nu)$
to be
\begin{equation}
E_g(\nu)=\frac{\hbar e|\Delta B|}{m^*(\nu)c}
\label{gapintro}
\end{equation}
where $m^*(\nu)$ is the effective mass calculated self--consistently at
an energy $E_g(\nu)$, and $\Delta B\equiv B-2\pi\tilde{\phi}{n}$ is
the deviation of the magnetic field from its value at $\nu=1/\tilde{\phi}$.
Here
we denote the average density of electrons by ${n}$, and we use units where
$\hbar=\frac{e}{c}=1$. Until now,
however, the conjecture (\ref{gapintro}) has not been checked by any explicit
calculation.

   In this paper we study both the even denominator  state and the way it is
   approached when the filling fraction is slightly away from it. We
   study the properties of the Fermi liquid formed at an
even denominator filling fraction (e.g., $1/2$), namely,
   the effective mass and the quasi--particle interaction function.
  We review the perturbative calculation leading to the divergence of the
  effective mass, and find that for the Coulomb case,
as $\omega\rightarrow 0$, the leading term
in the effective mass is,
\begin{equation}
m^*(\omega)=\frac{\tilde{\phi}^2}{2\pi}
\frac{\epsilon k_F}{e^2 }|\ln{\omega}|
\label{meff}
\end{equation}
where
$\epsilon$ is the background dielectric constant, $\omega$ is the
distance from the Fermi energy and the Fermi wavevector $k_F$ is related
to the electron density by $k_F=(4\pi n)^{1/2}$.
We then argue that Eq. (\ref{meff}) is the {\it exact} leading term in
the limit $\omega\rightarrow 0$.
This statement is based on the use of Ward's identity to account
 for the renormalization of the
  vertices in the diagram leading to the divergence of $m^*$. We directly
  calculate the energy gap at a filling factor of the form $\nu=\frac{p}
  {\tilde{\phi}p+1}$, and confirm the conjecture (\ref{gapintro}).
  Thus, for large $p$, the energy
gap we obtain is,
\begin{equation}
E_g\approx\frac{\pi \sqrt{2}}
{\tilde{\phi}^{3/2}}\frac{e^2}{\epsilon l_H}\frac{1}
{(\tilde{\phi}
p+1)\ln{(\tilde{\phi}p+1)}}
\label{gapp}
\end{equation}
where $l_H\equiv \sqrt{\frac{\hbar c}{eB}}$ is the magnetic length.
Following our discussion of the effective mass, we argue that our
  expression for the energy gap, Eq. (\ref{gapp}), is {\it exact}
 in the $p\rightarrow\infty$ limit \cite{Nummist}.
  In our study of the Landau quasi--particle interaction function
  at $\nu={1/\tilde{\phi}}$ we find
  that it includes a singular contribution that is just of the form necessary
  to explain the results of Kim {\it et.al.} \cite{Kim},
  i.e., of the form needed to
  cancel the effect of a diverging effective mass on the zero temperature
  long wavelength low frequency linear response function.
The expression we find for the Landau quasi--particle interaction function
is, however, not exact. In particular, our approximation does not properly
account for the effects of short wavelength and high frequency fluctuations.
These fluctuations are particularly important in the limit where the
electron's bare mass is vanishingly small, so that the cyclotron energy
becomes infinite and the actual electron states are projected onto
the lowest Landau level; our approximation incorrectly predicts the
magnitude of the Landau interaction function in this limit. The
short wavelength and high frequency fluctuations are not important
for the effective mass $m^*$ and the energy gaps at $p\rightarrow\infty$,
because these quantities are determined by infra--red divergences in the
self energy.

\begin{figure}[htbp]
\begin{center}
\leavevmode
\epsfbox{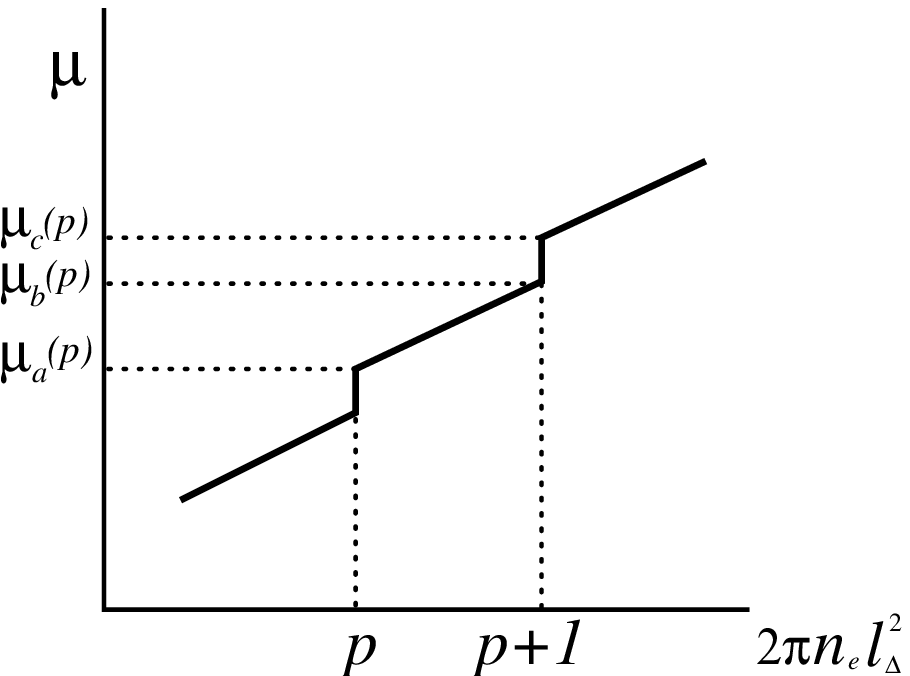}
\end{center}
\vskip 0.8truein
\caption{
 The variation of the chemical potential when the density is varied, and
$\Delta B$ is held fixed. The location of the origin and the units are
arbitrary. Discontinuous jumps take place when a vacant
Landau level starts being populated. Their magnitudes are the energy gaps
for the corresponding fractional quantum Hall states.  They reflect the
value of $m^*$.
Smooth variation takes place in
between discontinuous  jumps. The average slope of the curve is independent
of the singular part of $m^*$.}
\label{mufig}
\end{figure}

  Combining our analysis of the effective mass and the quasi--particle
  interaction function,
  we study the behavior of the chemical potential $\mu$ of the
composite fermions as
  the density ${n}$ is varied and $\Delta B$ is kept fixed. We find that the
  chemical potential jumps discontinuously whenever the density corresponds to
  an integer fermionic filling factor
  $p$. The magnitude of the jump is the energy gap corresponding
  to the fractional quantum Hall state at $\nu=\frac{p}{\tilde{\phi}p+1}$.
  As emphasized above, it reflects the divergence of $m^*$.
  However, we also find that the chemical potential varies continuously
  during the course of a population of a fermions' Landau level. This
  variation is such that the overall slope of $\mu({n})$ is independent of
 the divergent contribution to  $m^*$ (See Fig. (\ref{mufig})).

In our perturbative
study of the Fermion's effective mass, energy gaps and Landau
interaction function, we use the RPA expression for the gauge field propagator.
This use is then justified by our results. The singularities
we study result from the long wavelength low frequency behavior of the
gauge propagator. The latter are determined by the long wavelength low
frequency
limit of the fermion's polarization functions (the fermions irreducible
polarization bubble). As we show, and as was shown
before by several authors \cite{Kim}\cite{Aim}\cite{Wilczek2}, in that
limit, the fermions' irreducible polarization functions are not renormalized by
gauge field fluctuations.

  The outline of the paper is as follows: in Section (2) we construct a general
  framework for the discussion of the Fermi liquid formed by the composite
  fermions. In Section (3) we calculate perturbatively the parameters
  characterizing this Fermi liquid, and we discuss the regime of
validity of the
  perturbative approach. In Section (4) we examine the linear response
  function resulting from the parameters obtained in Sec. (3). In Section
  (5) we calculate the energy gaps for large $p$, and analyze the dependence of
  the chemical potential on the density for fixed $\Delta B$. In Section
(6) we comment on the effect of the coupling of the fermions to {\it
longitudinal}
gauge fluctuations. We conclude with a summary.

Several details of the analysis of Section (3) are given in Appendices
A and B.

\section{  General description of the Fermi liquid of the composite
fermions}

In this section we define the model to be considered, and we give a general
description of the Fermi liquid formed at even denominator filling fractions.
We consider electrons in a two dimensional electron gas, subject to a
magnetic field $B$. The electrons interact via a Coulomb interaction,
$V({\bf r}-{\bf r'})=\frac{e^2}{\epsilon|{\bf r}-{\bf r'}|}$. Using a
singular gauge transformation, the problem is mapped onto that of fermions
 subject to a magnetic field
$\Delta B= B-2\pi {n}\tilde{\phi}$.
(In our system of units, where  $\hbar=\frac{e}{c}=1$,
the flux quantum is $2\pi$.) In the case we discuss in this section, the
electronic filling factor is $1/\tilde{\phi}$, and then $\Delta B=0$ and the
fermionic filling factor, $p$,  is infinite.
In section (5) we discuss the case of finite $\Delta B$ and $p$. Generally,
the fermions' Hamiltonian is
\begin{equation}
\begin{array}{ll}
H=&{1\over {2m}}
\int d^2r\Psi^+({\bf r})[-i{\bf\nabla}
+\Delta{\bf { A}}({\bf r})-{\bf a}({\bf r})]^2\Psi({\bf r})
\\ \\
&+{1\over 2}
\int d^2r\int d^2r'[\Psi^+({\bf r})\Psi({\bf r})-{n}]V({\bf r}-{\bf r'})
[\Psi^+({\bf r'})\Psi({\bf r'})-{n}]
\\ \\
\end{array}
\label{ham}
\end{equation}
supplemented by the gauge condition ${\bf \nabla}\cdot{\bf a}=0 $
and by the constraint  ${\bf\nabla}\times{\bf a}({\bf r})=
2\pi\tilde{\phi}
[\Psi^+({\bf r})\Psi({\bf r})-{n}]$, where ${n}$ is the average
electron density.
In Eq. (\ref{ham}) $\Delta{\bf  A}({\bf r})=\frac{1}{2} \Delta B {\hat z}
\times{\bf r}$.
The constraint may be used to write the electrostatic term as
${1\over {2(2\pi\tilde{\phi})^2}}
\int d^2r\int d^2r' [{\bf\nabla}\times{\bf a}({\bf r})]
V({\bf r}-{\bf r'})
[{\bf\nabla}\times{\bf a}({\bf r'})]$.
The fermion's interaction with the transverse gauge field is then of the form
$-{\bf J}\cdot{\bf a}-{1\over{2m}}\rho {\bf a}^2$ where ${\bf J},\rho$ are the
fermions' current and density.  The Fermi
wave-vector of the fermions is $k_F=\sqrt{4\pi {n}}$.

At the mean field approximation, the Chern--Simons gauge field ${\bf a}$ is
replaced by its average value, and thus its interaction with the fermions
is neglected. In this section we attempt to describe the effect of these
interactions on the fermions within an extension of  Landau's
Fermi liquid theory. Landau's  theory describes the low energy
excitations of the system in terms of the occupation number function
$n({\bf k})$ \cite{Pita}--\cite{Nozieres}.
This function signifies the {\it difference}
 between the occupation
number of the state $\bf k$ at the excited state and the corresponding
number at the ground state. The energy density
corresponding to a given function
$n({\bf k})$ is
\begin{equation}
{\cal E}\Big(n({\bf k})\Big)={{\int\frac{d^2k}{(2\pi)^2} }} \epsilon_{\bf k}
n({\bf k})+
{1\over 2}{{\int\frac{d^2k}{(2\pi)^2} }}{{\int\frac{d^2k'}{(2\pi)^2} }}
 f({\bf k,k'})n({\bf k})n({\bf k'})
\label{landau}
\end{equation}
where $\epsilon_{\bf k}=
\frac{k^2}{2m^*}$ is the kinetic energy of a quasi--particle near the Fermi
level,
$m^*$ is its effective mass,
and $f({\bf k},{\bf k'})$ is the
Landau interaction function. Note that in the presence of a non--zero
vector potential, $\bf k$ is the dynamical momentum,
and not the canonical one.

The parameters $m^*, f$ include the entire
effect of the interaction near the Fermi level
in the case of  a neutral fluid such as $^3{\rm He}$ in its normal
state, but they conventionally
leave out the direct, Hartree, part of the interaction
in the case of electrons in a metal.
The difference between the two cases is
in the range of the interactions being short for Helium atoms and long
for electrons in metals. As developed in Silin's extension of Landau's
 theory \cite{Nozieres}, the
effect of long range interactions is taken into account
when linear response functions are calculated, by making
a distinction between the externally applied driving force and total
driving force. In the problem we
consider, the Chern--Simons interaction of the fermions with the gauge field
is long ranged. Consequently,
the direct Hartree part of both that interaction
and the electrostatic one should
be  separated from the rest. Thus, our construction of the Fermi liquid
picture for the composite fermions
starts by defining  the energy functional
(\ref{landau}) excluding the contribution of the direct Hartree part
of both interactions. Then,
 we use the equation of motion derived from
that functional to define and calculate a linear response function of
the fermions.

The equation of motion is conveniently written as an
integral equation for the function $\nu(\theta)$, defined by
$n({\bf k})\equiv\delta(k-k_F)\nu(\theta)$, where $\theta$ is the angle between
$\bf k$ and some reference direction
\cite{Pita}--\cite{Nozieres}. The function $\nu(\theta)$
describes the deformation of the Fermi surface in the direction
$\theta$. The equation of motion for $\nu(\theta)$
is
\begin{equation}
(\omega-{ q}{ v_F^*}\cos{\theta})\nu_{{\bf q},\omega}
(\theta)-\frac{1}{(2\pi)^2}{ q}k_F\cos{\theta}
\int d\theta' f(\theta-\theta')\nu_{{\bf q},\omega}(\theta')=
{\bf\hat k}\cdot {\bf F}_{{\bf q},\omega}
\label{eomnu}
\end{equation}
where
the angle $\theta$ is the angle between the vectors ${\bf k}$ and ${\bf q}$;
the function
$f(\theta-\theta')$ is the quasi--particle interaction function $f({\bf k,k'})$
for two wave-vectors at the Fermi surface, at angles $\theta,\theta'$ relative
to $\bf q$;
$\bf F$ is the driving force, characterized by a wave--vector $\bf q$
and a frequency $\omega$; and the renormalized Fermi velocity is denoted by
$v_F^*$.
Eq. (\ref{eomnu}) is valid for $q\ll k_F$ and $\omega\ll\mu$.

In Landau's Fermi
liquid theory the linear response function is extracted from Eq. (\ref{eomnu}).
This equation  relates the
deformation function $\nu(\theta)$
to the driving force $\bf F$.  In terms of $\nu(\theta)$, the
fermions' density is
$\rho_{{\bf q},\omega}=k_F\int d\theta \nu_{{\bf q},\omega}(\theta)$
while the current density
is  ${\bf J}_{{\bf q},\omega}=\int d^2k \delta(k-k_F)
{\bf u}_{\bf k}
\nu_{{\bf q},\omega}(\frac{\bf k\cdot q}{kq})$, with ${\bf u_k}\equiv
\frac{\bf k}{m^*}+\int d^2k' f({\bf k},{\bf k'})\delta(\epsilon_{\bf k'}-\mu)
\frac{\bf k'}{m^*}$.
The driving force can be expressed in terms of derivatives of
a three--vector
potential $A_\nu$, as ${\bf F}_{{\bf q},\omega}=-
i\omega {\bf A}_{{\bf q},\omega}+i{\bf q}A_{0,{\bf q},\omega}$.
Thus, by means of Eq. (\ref{eomnu}), the current vector
$J_\mu$ can be related to the vector potential $A_\nu$
by $J_\mu=\Pi_{\mu\nu}A_\nu$,  and the linear
response function $\Pi_{\mu\nu}$
 can be calculated. Generally, the linear response
 function is a matrix relating the current--density three--vector $J_\mu$ to
 the three--vector potential $A_\nu$.
Since the three--vector $J_\mu$ is
constrained by the conservation of charge and the three--vector
$A_\nu$ is constrained  by a gauge condition ${\bf \nabla\cdot A}=0$, the
linear response function  can in fact be described by
 a  $2\times 2$ matrix $\Pi_{\mu\nu}({\bf q},\omega)$, with the two
indices taking the values
 $0$ (for the time component), and $1$ (for the transverse
component).

Now we turn to discuss  the effect of the Hartree term.
The function $\Pi$ relates the current $J_\mu$
to the {\it total} vector potential acting on the fermions. This
vector potential is composed of an externally applied part and a
part induced by the fermion current and density. A fermion
current--density vector $J_\mu$ induces a vector potential
$A^{ind}_\nu=V_{\mu\nu}J_\mu$ where the matrix $V$ is given by,
\begin{equation}
V=\left (
\begin{array}{ll}
{{2\pi e^2}\over{\epsilon q}} & \frac{2\pi i\tilde{\phi}}{q} \\
      -\frac{2\pi i\tilde{\phi}}{q} & 0\\
        \end{array}\right )
\label{induced}
\end{equation}
The diagonal term is the electrostatic potential induced by a charge
density. The off diagonal terms
describe the vector potential induced by the fictitious flux tubes attached
to the fermions.

\begin{figure}[htbp]
\begin{center}
\leavevmode
\epsfbox{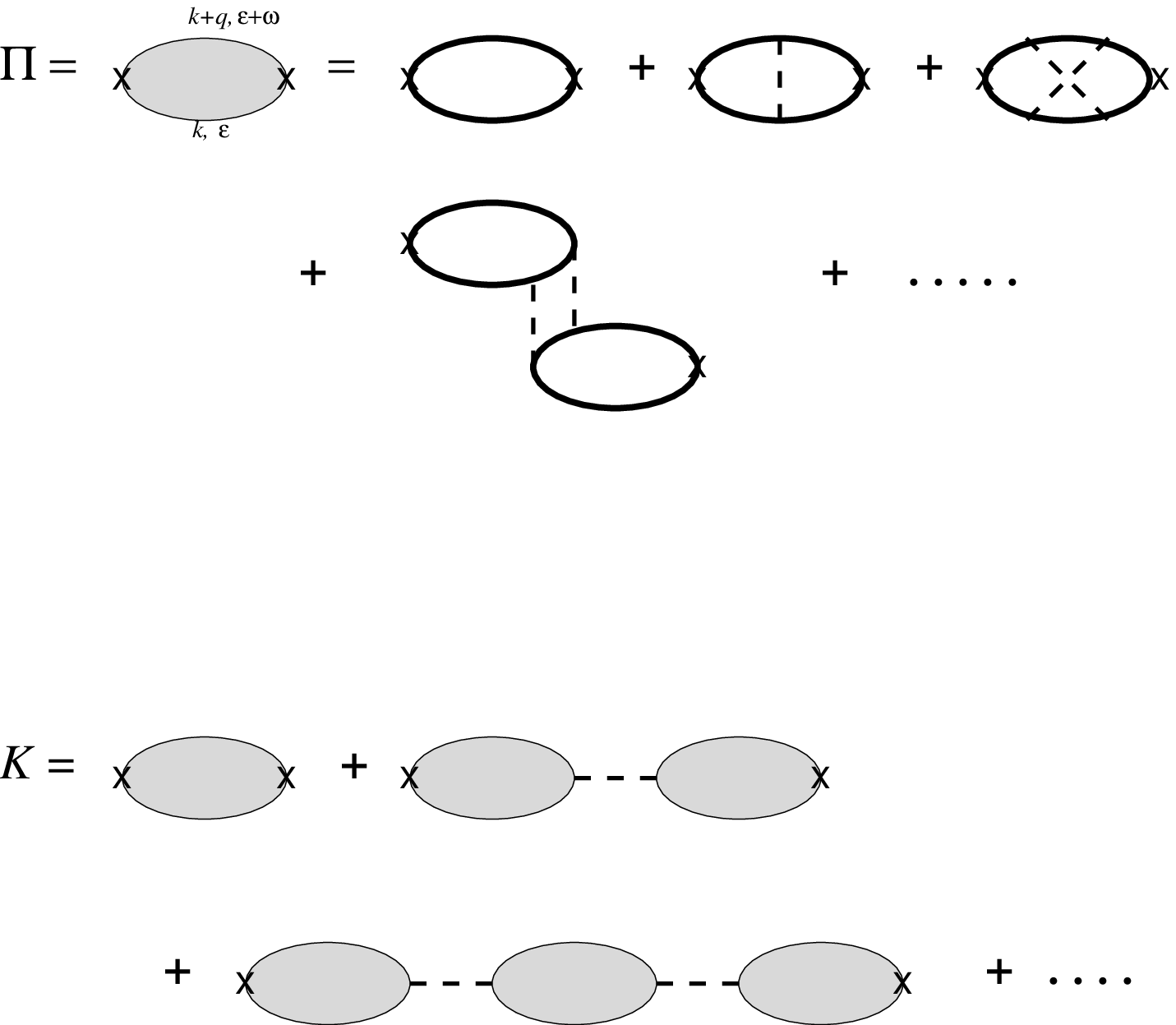}
\end{center}
\vskip 0.8truein
\caption{
The diagrammatic representation of the linear response functions $\Pi$
and $K$. The response function
$\Pi$ is the fully dressed
fermion--hole bubble, irreducible with respect to a single
gauge field line. The response function $K$ is the geometric series
summing strings of bubbles connected by single gauge field propagator line.
Solid lines are fermion propagators, while dashed lines are gauge field
propagators.
The physical meaning of both is discussed in the text.}
\label{kpifig}
\end{figure}

We can now define
the matrix $K$, relating the current--density vector to the externally applied
vector potential $A^{ext}_\nu$ by  $J_\mu=K_{\mu\nu}A^{ext}_\nu$.
Since $A_\nu=A_\nu^{ext}+A_\nu^{ind}$,
$K$ is related to $\Pi$ and $V$ by
\begin{equation}
K^{-1}=V+\Pi^{-1}
\label{kpi}
\end{equation}
Thus, the matrix $K$ describes the response of the system to an externally
applied vector potential, including both the Hartree and non--Hartree parts
of the interaction. In terms of a diagrammatic expansion, the function $\Pi$
is the fermion--hole proper
polarization part (irreducible with respect to a single $V$ line),
while the function $K$ is a sum of a
geometric series, i.e., of a chain of  proper polarization parts
connected by single interaction
lines $V$ (see Fig. (\ref{kpifig})).

The difference between the functions $\Pi$ and $K$ should be particularly
stressed in the context of the
compressibility. Generally, the compressibility is given by the
$\omega=0, q\rightarrow 0$ limit of the density linear response
function.  Denoting the ground state energy per unit volume
by $\epsilon_0(n)$ where $n$ is the density of particles, the inverse
compressibility is
$\frac{1}{\chi}\equiv n^2\frac{\partial^2\epsilon_0}{\partial n^2}$
\cite{Nozieres}.
If the derivatives are taken with the external magnetic field $B$ held fixed,
 the change in
density induces a uniform magnetic field $\Delta B$ on the fermions. The
fermions respond to
 the combination of the driving force
and the induced uniform $\Delta B$, and their response is
therefore  described by the linear
response function $K$. The function $\Pi$  describes the case in which a
change in the density is accompanied by a change in the external magnetic field
$B$, in such a way that $\Delta B$ is kept constant. The two conditions
yield two very different compressibilities, as can be seen by
analyzing the case of non--interacting electrons around $\nu=1/2$.
If the density is changed
around $\nu=1/2$ with $B$ being kept constant, $\frac{1}{\chi}=0$.
However, if $\Delta B$ is to be kept constant, the variation
of the density results in a variation of the cyclotron frequency,
and thus a variation of the kinetic energy of the
lowest Landau level. Then,
$\frac{1}{\chi}=\frac{\pi\tilde{\phi}}{m}$. We come back to this
distinction in Sec. (5), where we discuss the gaps in the fractional quantum
Hall states.

Another use of  Eq. (\ref{eomnu}) is for  the analysis of the excitation modes
of the system. The solutions of the equation in the absence of $\bf F$
are poles of $\Pi$. They are characterized
by a  phase velocity $\omega/q$ and by an eigenfunction $\nu(\theta)$.
Those solutions for which $\rho_{{\bf q},\omega}={\bf J}_{{\bf q},\omega}=0$
are unaffected by the matrix $V$. They are poles of $K$ as well.
Solutions that do not satisfy these requirements are not poles of
$K$.
 Some insight into
the solutions of Eq. (\ref{eomnu}) in the absence of a driving force
 is obtained by describing $f(\theta)$ by its Fourier components,
 the  "Landau parameters"
\begin{equation}
f_l\equiv \frac{1}{2\pi}
\int_0^{2\pi} d\theta f(\theta)e^{il\theta}
\label{param}
\end{equation}
 In "well behaved"
Fermi liquids, the ratio $m^*/m$ is a finite number, and only a small, finite
number of Landau parameters are appreciably different from zero. In that case
Eq. (\ref{eomnu}) has a continuum of excitation modes at $\omega<v_F^* q$,
and a small number of discrete states at $\omega>v_F^* q$. As we show in
the next sections, the Fermi liquid we consider does not conform to this
description.

To conclude this section, we emphasize again the main points in the
Fermi liquid picture we construct for the composite fermions. The
energy of an excited state is described by the functional (\ref{landau}).
The parameters in this functional, $m^*$ and $f$, will be calculated in
the next section. The Landau function $f$ does not include the Hartree
contribution of either the Coulomb or the Chern--Simons interaction.

The linear
response of the system can be described in terms of two types of
response functions. The
first, $\Pi_{\mu\nu}({\bf q},\omega)$,
describes the response of the function to the total driving
force, while the second,
$K_{\mu\nu}({\bf q},\omega)$, describes the response of the system to
the externally applied driving force. In the limit $\omega=0$ and
$q\rightarrow 0$,  $\Pi$ describes the static compressibility when
$\Delta B$ is kept fixed,  while $K$ describes the compressibility
when  $B$
is kept fixed. The poles of the linear response functions can be
found, within Fermi liquid theory,
 by use of the equation of motion for $n({\bf k})$, Eq. (\ref{eomnu}).

\section{Perturbative calculation of the Fermi liquid parameters}

\subsection{Starting Point}
In this section  we calculate perturbatively the parameters
characterizing the Fermi liquid state at $\Delta B=0$, namely $m^*$
and $f$.
Landau's theory of the Fermi liquid expresses  $m^*$ and
$f$ in terms of the self energy and the interaction operator of fermions at
the Fermi level. The latter, in turn, are expressed in terms of the
propagators for the fermions and the gauge field. We start this section by
writing the  propagators of the fermions and the gauge field. Then, we
use these propagators to calculate the self energy, and use the self
energy to show that the effective mass diverges. Then, we turn to
discuss the Landau function, and pay a special attention to keeping
internal consistencies between the approximations used. The Landau
function we find turns out to have unique properties, which we discuss in
detail.

The fermion propagator in space--time representation is defined, as usual, as
\begin{equation}
{\tilde G}({\bf r},t;{\bf r'},t')\equiv
-i\langle 0|{\hat T}\psi^+({\bf r},t)\psi({\bf r'},t')|0\rangle
\label{gfdef}
\end{equation}
where $|0\rangle$ is the ground state, ${\hat T}$ is the time ordering
operator and the creation and destruction operators are expressed in
the Heisenberg picture. The gauge field propagator is defined
in a similar way. For the case of $\Delta B=0$, both the Hamiltonian
(\ref{ham}) governing the time evolution of
the system and its ground state are invariant to space and time translations.
Therefore, ${\tilde G}$ is a function of ${\bf r-r'}$ and $t-t'$ only,
and it is diagonal in Fourier energy--momentum representation. The same
applies also to the gauge field propagator, and, obviously, to the bare
propagators.

The bare propagator for
the fermions is, for $\Delta B=0$,
\begin{equation}
G_f^{(0)}( {\bf k},\omega)=\frac{1}{\omega+\mu-{{k^2}\over{2m}}+
i\eta{\rm sgn}(\omega)}  \ \
\label{fermiongf}
\end{equation}
where  $\mu$ is the chemical potential for the transformed fermions,
and $\bf k$ is a wavevector.

Since ${\bf \nabla}\cdot{\bf a}=0$ the gauge field propagator is a $2\times 2$
matrix $D_{\mu\nu}$, with $\mu=0$ denoting the time component and
$\mu=1$ denoting the transverse spatial component.
The gauge field propagator we use in the following calculation is obtained
by the random phase approximation (RPA).
Its value for
$\nu=1/\tilde{\phi}$ ($p=\infty$) was calculated in
Ref. \cite{Hlr}, and was found to be
\begin{equation}
D({\bf q},\omega)=\left(
\begin{array}{ll}
{m\over{2\pi}} \ \ &\ \  -i{q\over{2\pi\tilde{\phi}}}\\
i{q\over{2\pi\tilde{\phi}}} \ \ &\ \  i\frac{2{n}}{k_F}\frac{\omega}{q}-
q^2({1\over{12\pi m}}+{{e^2}\over{2\pi\tilde{\phi}^2\epsilon q}})
\end{array} \right)^{-1}
\label{gaugegf}
\end{equation}

In terms of the self
energy of states
at the Fermi level
the effective mass $m^*$ is \cite{Mahan}:
\begin{equation}
m^*=m\frac{1-\frac{\partial\Sigma}{\partial\omega}|_{\omega=0}}
          {1+\frac{\partial\Sigma}{\partial\xi_k}|_{k=k_F}}
                  \label{effmass}
\end{equation}
where $\xi_k=\frac{k^2}{2m}-\mu$.

\subsection{Lowest--Order--In--$D$ approximation}
\subsubsection{Fermion Effective Mass}

\begin{figure}[htbp]
\begin{center}
\leavevmode
\epsfbox{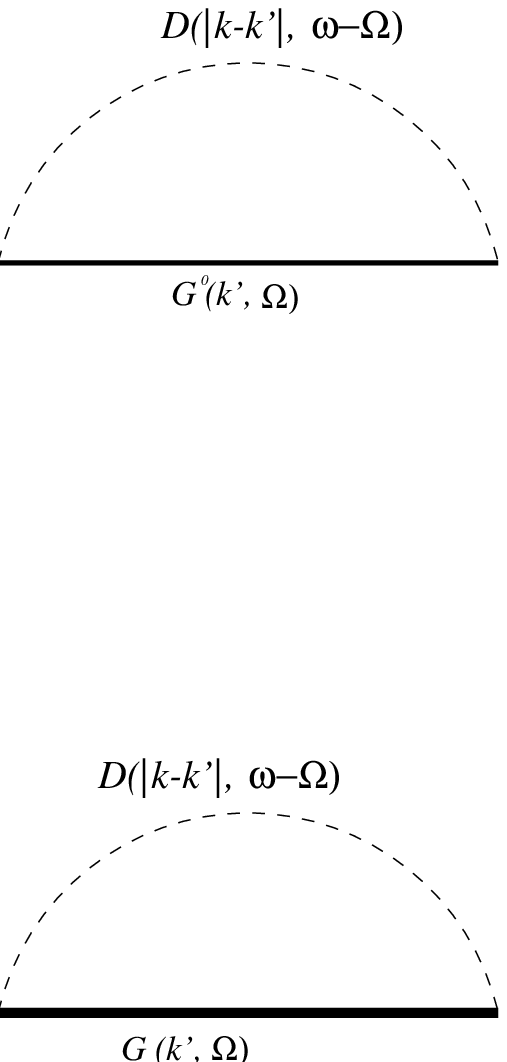}
\end{center}
\vskip 0.8truein
\caption{}{
The self energy diagrams. (a.) The lowest order in $D$ diagram, leading to
Eq. (\ref{sezero}). The dashed line represents the RPA expression for the gauge
field propagator (Eq. (\ref{gaugegf}).
(b.) The self consistent approximation, leading to Eq.
(\ref{sesc}).
}
\label{sigmafig}
\end{figure}

The self energy of the fermions
is given, to first order in the gauge field propagator
$D$, by the diagram given in Fig. (\ref{sigmafig}a),
\begin{equation}
\Sigma(k,\omega)=i\int\frac{d^2k'}{(2\pi)^2}\int \frac{d\Omega}{2\pi}
\frac{1}{m^2}\frac{({\bf k}\times{\bf k'})^2}
       {|{\bf k}-{\bf k'}|^2}
D(|{\bf k}-{\bf k'}|,\omega-\Omega)
G_f^0(k',\Omega)
\label{sezero}
\end{equation}
There are two singularities in the evaluation of the integral (\ref{sezero}).
The first results from the high frequency regime of
the longitudinal gauge propagator $D_{00}$. It does not
affect the effective mass, and its discussion is therefore deferred to
section (6). The second results from the infrared regime of $D_{11}$, and
is the source of the logarithmic divergence of the effective mass, as we
now turn to discuss.
In the long wavelength low frequency limit $D$ is dominated by its
diagonal transverse component, given by
 $D_{11}(q,\omega)\approx[i\frac{2{n}}{k_F}\frac{\omega}{q}-
q{{e^2}\over{2\pi\tilde{\phi}^2\epsilon}}]^{-1}$. We focus now
on that element of $D$, and
neglect all the others.
The evaluation of the integrals in (\ref{sezero})
is similar to that carried out in studies
of the electron--phonon coupling \cite{Pita}. It is
convenient to evaluate $\Sigma(k,\omega)-\Sigma(k,0)$ and to introduce a
variable $q=|{\bf k}-{\bf k'}|$. Most of the contribution to the
$k'$ integral comes from $k'\approx k_F$. From the form of
the gauge field propagator we observe that the important contribution to
the $q$ integral comes from
$q>2\pi\tilde{\phi}\sqrt{\frac{2 {n}\epsilon\omega}{e^2 k_F}}$.
The integral measure is transformed to
\begin{equation}
\int_0^\infty k' dk'\int_0^{2\pi}d\theta\longrightarrow
2\int_0^\infty dk'\int_{|k-k'|}^{k+k'}
dq \frac{2k'q}{\sqrt{q^2-(k-k')^2}
\sqrt{(k+k')^2-q^2}}
\label{measure}
\end{equation}

It is convenient to consider  two regimes. In the first regime
 $\omega>\frac{e^2k_F}{2\pi\tilde{\phi}^2\epsilon}
\frac{(k-k_F)^2}{k_F^2}$, the integral measure can be approximated by
$ 2\int_0^{\infty}dk'\int_0^{2k_F} dq$.
Then, the integral over $k'$ restricts $\Omega$ to satisfy $0<\Omega<\omega$,
and the self energy becomes
\begin{equation}
\Sigma(k,\omega)=\frac{\tilde{\phi}^2}{2\pi}
\frac{\epsilon k_F}{me^2}
\omega\ln{\left (\frac{4e^2k_F}{\tilde{\phi}^2\epsilon\omega}\right )}
+i\frac{\tilde{\phi}^2}{4}
\frac{\epsilon k_F}{me^2}
\omega
\label{sigexp}
\end{equation}
Near the Fermi level the real part of $\Sigma$ is larger than its imaginary
part, thus justifying the notion of a quasi--particle.
 The origin of
the logarithmic singularity is
the
$1\over q$ divergence of $D$ for $\omega=0$.
Note that the important
contributions to $\Sigma$ come from a limited range of frequencies
$0<\Omega<\omega$, but
from a large range of wave-vectors
$2\pi\tilde{\phi}\sqrt{\frac{2 {n}\epsilon\omega}{e^2 k_F}}<q<2k_F$.
Within the approximation used, $\Sigma$ is independent of $k$.
Corrections to that approximation (e.g., to the approximation of the
integral's measure) yield a weak and regular dependence on $k$.
Substituting the approximation (\ref{sigexp}) in Eq. (\ref{effmass})
we get a logarithmically diverging effective mass, Eq. (\ref{meff}).
Note also that the residue
$(1-\frac{\partial \Sigma}{\partial\omega})^{-1}$ vanishes
at the Fermi level.

In the
second regime $\omega<\frac{e^2k_F}{2\pi\tilde{\phi}^2\epsilon}
\frac{(k-k_F)^2}{k_F^2}$,
the integral measure (\ref{measure})
is multiplied by a factor of $q\over {k-k_F}$
and the singularity in the self energy is weaker.
Near the Fermi surface the mass shell
is defined by $\omega\ln{\omega}\propto |k-k_F|$.
Thus,
 the first regime determines the
quasi--particles dispersion relation and  the effective mass is infinite.
Later we shall argue that the above conclusions  do not change
when the set of diagrams included in the approximation
for the self energy is extended.

\subsubsection{Landau Interaction Function}

\begin{figure}[htbp]
\begin{center}
\leavevmode
\epsfbox{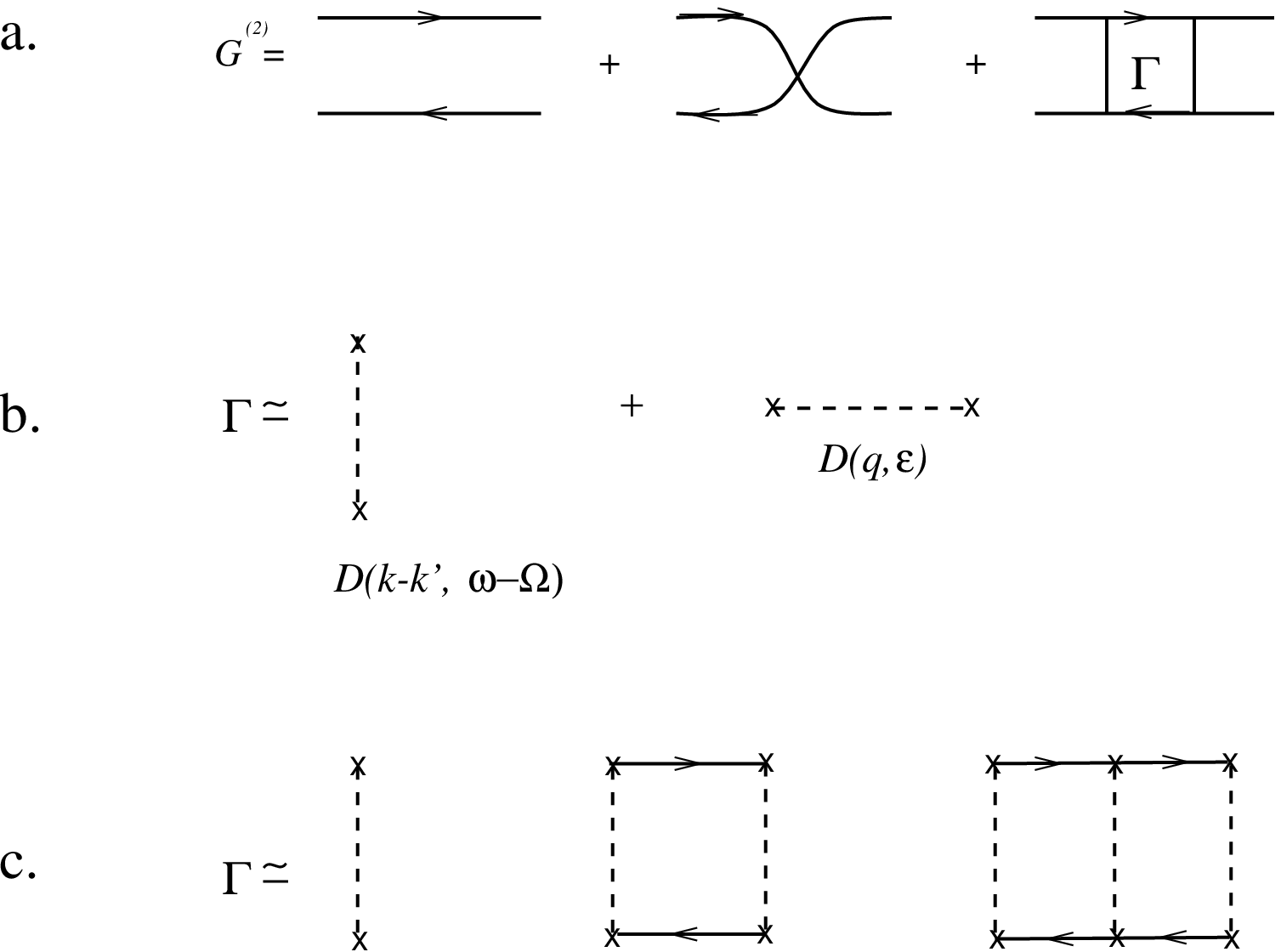}
\end{center}
\vskip 0.8truein
\caption{}{ The proper interaction operator ${\tilde\Gamma}$
and the approximations we employ for it. (a.) The
interaction operator $\Gamma$ is defined via the fermions'
two particle Green function $G^{(2)}$.
In this figure, solid lines represent exact single particle Green's functions.
(b.) The simplest approximation for $\Gamma$, used in calculating Eq.
(\ref{zerogamma}). The dashed lines represent the RPA expression for the
gauge field propagator (Eq. (\ref{gaugegf})),
and the crosses represent bare vertices. The left diagram is proper, and is
included in $\tilde\Gamma$. The right diagram is improper, and is not
included in $\tilde\Gamma$.
(c.) The ladder diagrams sum for ${\tilde\Gamma}$, summed in
 Appendix B.}
\label{gammafig}
\end{figure}

We now turn to discuss the Landau function $f$.
For a rotationally invariant system the
Landau function $f({\bf k},{\bf k'})$ is a function of $\theta$,
the angle between the vectors ${\bf k},{\bf k'}$ (note that $k\approx k'
\approx  k_F$).
Standard Fermi liquid theory expresses the Landau function in terms of the
Green's function residue $z$ and the proper part of the interaction operator
$\tilde{\Gamma}({\bf k}_1,\omega_1;{\bf k}_2,\omega_2;{\bf k}_1+{\bf q},
\omega_1+\omega; {\bf k}_2-{\bf q},
\omega_2-\omega)$   as \cite{Pita},
\begin{equation}
f({\bf k},{\bf k'})= z_{k_F}^2{\rm lim}_{\omega\rightarrow 0}
{\tilde\Gamma}^\epsilon({\bf k},0;{\bf k'}0;{\bf k'},
\omega; {\bf k},-\omega)
\label{fgamma}
\end{equation}
where $z_{k_F}$ is the Green's function residue for poles on the Fermi surface,
and $
{\tilde\Gamma}^\epsilon({\bf k},\omega;{\bf k'}\Omega;{\bf k'},
\Omega; {\bf k},\omega) \equiv
{\rm lim}_{\epsilon\rightarrow 0}{\rm lim}_{q\rightarrow 0}
{\tilde\Gamma}({\bf k},\omega;{\bf k'}\Omega;{\bf k'+q},
\Omega+\epsilon; {\bf k-q},\omega-\epsilon)$.
The full interaction operator
${\Gamma}({\bf k},\omega;{\bf k'}\Omega;{\bf k'+q},
\Omega+\epsilon; {\bf k-q},\omega-\epsilon)$,
including both the proper and
improper parts, is defined through the two particle Green's function, in
a way described in Fig.  (\ref{gammafig}). The proper part,
$\tilde\Gamma$, is the sum of all diagrams of $\Gamma$ that are
irreducible with respect to cutting a single interaction line, {\it
carrying a momentum $q$}.
The exclusion of the improper diagrams from the Landau function stems from
the distinction between the linear response function $\Pi$ and $K$,
defined in the previous section. The Landau function determines the
linear response function $\Pi$, through Eq. (\ref{eomnu}). Thus, it should
include only diagrams that are "building blocks" for
building  the particle--hole proper
polarization part (see Fig. (\ref{kpifig})). The diagrams excluded from
the Landau function are those that serve to build the linear response function
$K$. Examples are given in Fig. (\ref{sigmafig}). A comprehensive discussion
of the distinction between $\Gamma$ and $\tilde\Gamma$ is given in Ref.
\cite{Nozieres}.

For our purpose, we find it convenient to rewrite
Eq. (\ref{fgamma}) in a slightly different form. We define
\begin{equation}
\begin{array}{ll}
\left\{G^2({\bf k},\omega)\right\}_\epsilon\equiv
{\rm lim}_{\epsilon\rightarrow 0}
G_f({\bf k},\omega+\epsilon)G_f({\bf k},\omega) \\ \\
\left\{G^2({\bf k},\omega)\right\}_{\bf q}
\equiv{\rm lim}_{{\bf q}\rightarrow 0}
G_f({\bf k+q},\omega)G_f({\bf k},\omega)
\end{array}
\label{gfsq}
\end{equation}
Since $\left\{G^2({\bf k},\omega)\right\}_\epsilon-
\left\{G^2({\bf k},\omega)\right\}_{\bf q}=\frac{2\pi i m^*z_{k_F}^2}{k_F}
\delta(\omega)\delta(k-k_F)$, \cite{Pita}--\cite{Nozieres},
 Eq. (\ref{fgamma}) can be
written as
\begin{equation}
f({\bf k},{\bf k'})={\rm lim}_{\omega\rightarrow 0}
{{-i(2\pi)^2}\over {m^*(\omega)}}
\int \frac{k'dk' }{(2\pi)^2} \int \frac{d\Omega}{2\pi}
{\tilde\Gamma}^{\epsilon}({\bf k},\omega;{\bf k'}\Omega;{\bf k'},
\Omega; {\bf k},\omega)
\left[\{G^2({\bf k},\Omega)\}_\epsilon-
\{G^2({\bf k},\Omega)\}_{\bf q}\right ]
\label{fgammab}
\end{equation}

The extraction of an approximation for the Landau function from
 an approximation for the interaction operator and
the Green's function residue should be done carefully, since
 the approximations for $z_{k_F}$  and for $m^*$ do not result from
a systematic expansion in powers of a small parameter. Consistency between
the approximations used is established by verifying that they satisfy the
symmetries of the problem, namely, gauge invariance and galilean
invariance. To be consistent
with both symmetries, the approximations used
should satisfy Ward's identities. Gauge invariance
dictates the identity \cite{Pita},
\begin{equation}
\frac{\partial\Sigma}{\partial\omega}=
i\int \frac{d^2k'}{(2\pi)^2}  \int \frac{d\Omega}{2\pi}
{\tilde\Gamma}^\epsilon({\bf k},\omega;{\bf k'}\Omega;{\bf k'},
\Omega; {\bf k},\omega)G_f^2({\bf k'},\Omega)
\label{ward}
\end{equation}
(Generally, Ward's identities relate $\Sigma$ to $\Gamma$, and not to
$\tilde\Gamma$. However, since here there is no direct contribution to
the self energy,  the contribution of the improper diagrams of
$\Gamma$ to the right hand side of Eq. (\ref{ward}) vanishes).
When the self energy is evaluated by Eq. (\ref{sezero}),
Ward's identity (\ref{ward})
suggests the approximation of the interaction operator by
\begin{equation}
{\tilde\Gamma}^\epsilon({\bf k},\omega;{\bf k'}\Omega;{\bf k'},
\Omega; {\bf k},\omega)=\frac{1}{m^2} \frac{({\bf k}\times{\bf k'})^2}
                                {|{\bf k}-{\bf k'}|^2}
                                D(|{\bf k}-{\bf k'}|,\omega-\Omega)
\label{zerogamma}
\end{equation}
and of the Green's function by $G_f=G_f^{(0)}$. It can be shown that this
approximation is consistent with the other Ward identities, too.

Ward's identities are a test of consistency of the approximations for $\Sigma$
and $\tilde\Gamma$.
Attempting to use the consistent approximations for $\Sigma$ and
${\tilde\Gamma}$ to formulate consistent
approximations for the effective mass and the
Landau function, we are in need for a similar consistency test. This test is
given by the Fermi liquid identities,
\begin{equation}
\begin{array}{cc}
\frac{d\mu}{dn}=\frac{2\pi}{m^*}+f_0\ , \\
\frac{1}{m}=\frac{1}{m^*}+\frac{f_1}{2\pi}\ .
\end{array}
\label{fltwi}
\end{equation}
It can be shown that if  the
approximations for $\Gamma$, $G_f$  and $\Sigma$
used in Eq. (\ref{fgammab}) (where the approximation for $\Sigma$ is
used to determine $m^*$) satisfy the Ward identity (\ref{ward}), the identities
(\ref{fltwi}) are satisfied, too.
We show that for the identity involving $f_0$
in Appendix A.

Consequently, in using Eq. (\ref{fgammab}) to
consistently  approximate  the
Landau function, the Green's function would be approximated by $G_f^0$,
the interaction operator by Eq. (\ref{zerogamma}), and the effective mass
evaluated using the self energy (\ref{sigexp}).
 Due to the divergence of $m^*$ at the Fermi energy, the limit
$\omega\rightarrow 0$ of Eq. (\ref{fgammab}) should be taken carefully. Keeping
$\omega$ on the right hand side of (\ref{fgammab}) small but finite, we may
define a frequency dependent Landau function, $f_\omega(\theta)$.
Using Eq. (\ref{sezero}) for the self energy, Eq. (\ref{zerogamma})
for the interaction operator and $G_f^{(0)}$ for the Green function
we find
\begin{equation}
f_\omega(\theta)=\frac{2\pi k_F^2}{m m^*(\omega)}\cos^2{\theta\over 2}\
D(2k_F\sin\frac{\theta}{2},\omega)
\label{ftheta}
\end{equation}
For small $\omega$,  $f_\omega(\theta)$ is strongly peaked around
$\theta=0$, and
the number of appreciably non--zero Landau parameters is very large. At  the
limit of zero frequency,
\begin{equation}
{\rm lim}_{\omega\rightarrow 0}f_\omega(\theta)=
\frac{(2\pi)^2}{m}\delta(\theta)
\label{fdelta}
\end{equation}
and all Landau parameters are equal.
The finite frequency $m^*(\omega), f_\omega(\theta)$ should be used when
the equation of motion (\ref{eomnu}) is analyzed. We present a partial
analysis of this equation in the next section, and defer part of it to a
future publication.

\subsection{Self Consistent Green's Function Approximation and Beyond}

 The remainder of this section is
devoted to an extension  of the set of diagrams included in our approximation
for $\Sigma$ and ${\tilde\Gamma}$. This extension suggests the conclusion that
our approximation for $m^*(\omega)$, obtained from Eq. (\ref{sezero}), is
exact in the limit $\omega\rightarrow 0$, while our approximations for
the Landau functions are not.

The approximation we have used so far  can be improved by evaluating the
self energy self--consistently, i.e., by solving the equation,
\begin{equation}
\Sigma(k,\omega)=
\int\frac{d^2k'}{(2\pi)^2}\int\frac{ d\Omega}{2\pi}
 \frac{1}{\Omega-\epsilon(k')-
\Sigma(k',\Omega)}
\frac{1}{m^2}\frac{({\bf k}\times{\bf k'})^2}
       {|{\bf k}-{\bf k'}|^2}
D_{11}(|{\bf k}-{\bf k'}|,
\omega-\Omega)
\label{sesc}
\end{equation}
The self energy obtained by Eq. (\ref{sesc}) is a sum of all "rainbow"
diagrams.
The singular parts of the effective mass and the Landau function
are unaffected by this improvement of the approximation.
To see that, follow the same route taken in
evaluating Eq. (\ref{sezero}). The integral over $k'$ is proportional to
the ratio $\frac{z_{k_F}m^*}{k_F}=\frac{1}{v_F}$. The residue is
vanishingly small and the mass is infinitely large, but their product is
finite, and is unaffected by the perturbation. Carrying out the rest of
the integrals, we then find that the singular part of the self energy
obtained from Eq. (\ref{sesc}) is identical to that obtained from
(\ref{sezero}).
 Consequently, so is the effective
mass. As for the Landau function, by using the relation between
the bare mass, the effective mass and $f_1$ (Eq. (\ref{fltwi})),
 it is easy to see that
$f_1$ is unaffected, too. The same can be shown for $f(\theta)$:
when the self energy is approximated by the self consistent
expression (\ref{sesc}), Ward's identity requires the
approximation of the interaction operator by a sum of  ladder diagrams
(see Fig. (\ref{gammafig}))
and the Green function by the self consistent solution of Eq.
(\ref{sesc}),
 $G_f\approx{[\Omega-\epsilon(k')-
\Sigma(k',\Omega)]^{-1}}$. When these values of ${\tilde\Gamma}$ and $G_f$ are
substituted in Eq. (\ref{fgammab}), $f(\theta)$ is found to be
identical to that given in Eq. (\ref{ftheta}). We carry out this calculation
in Appendix B.

A further improvement of the approximation (\ref{sezero}) for the self
energy might, in principle, be achieved by replacing the interaction vertices
by  dressed ones.  As far as the effective mass is concerned, this replacement
may lead to a modification of the logarithmic singularity
in the limit $\omega\rightarrow 0$, to a modification of the prefactor of the
logarithm or to no modification of this singularity. We now argue that the
last is true, i.e., that {\it the effective mass (\ref{meff}) is exact in the
limit $\omega\rightarrow 0$.} The essence of our argument is the
observation, discussed below Eq. (\ref{sigexp}),
 that for a state with an energy $\omega$ close to the Fermi
level, the contributions to the self energy come from small energy
transfers $0<\Omega<\omega$, but from a wide range of momentum transfers
$q_0<q<2k_F$,
where $q_0\propto\sqrt\omega$.  The logarithmic
divergence of the effective mass is a result of the long wavelength
contribution, of order of $\sqrt\omega$. Thus, the interaction vertex of
significance is that in which both the momentum $q$ and the energy
$\omega$ transferred to the gauge
field approach zero, with the ratio $\omega/q\sim \sqrt{\omega}\rightarrow 0$.

 As a first step in our
argument, consider separating the gauge field propagator $D$ into
long wavelengths (small $q$) and short wavelengths (large $q$)
parts. The divergence of the effective
mass results from the long wavelength part. Now, imagine repeating the
calculation of the self energy leading to Eq. (\ref{sezero}) or (\ref{sesc}),
but with the fermions' Green's functions, the gauge field propagator and the
interaction vertices  renormalized by the short wavelength gauge
fluctuations. Moreover, imagine that this renormalization is done exactly,
i.e., that $G_f$, $D$ and the vertices are dressed by all possible lines of
short wavelengths interactions (see Fig. (\ref{vertexfig}) for some low
order contributions for the dressing of the vertex).
What would be the effect of this renormalization on the
logarithmic singularity in the self energy? Let us see how it affects
each of the ingredients of the diagram in Fig. (\ref{vertexfig}). First,
consider the fermion's
Green's function. The interaction of
the fermions' with the short wavelength gauge fluctuation induces a
regular self energy  in the fermions' Green function.
We denote that self energy by $\Sigma^>(k, \omega)$.
Second, consider the vertices. The self energy $\Sigma^>(k,\omega)$
determines not only the renormalization of the Green's function, but also that
of the vertices.
 The latter is determined by a Ward identity:
in the limit $\omega/q\rightarrow 0$, the interaction vertex is multiplied
by $1+\frac{\partial\Sigma^>}{\partial\epsilon_k}$
\cite{Nozieres}. Lastly,
the gauge field propagator is not likely to be renormalized by the short
wavelength gauge fluctuations: it is determined by the electron density
and the electron's charge, none of which is subject to
renormalization.

Armed with these
observations, we may now calculate the self energy due to the long
wavelengths fluctuations, with the short wavelengths fluctuations taken into
account exactly. The diagrams in Fig. (\ref{sigmafig})
 include two vertices, and each is
multiplied by $1+\frac{\partial\Sigma^>}{\partial\epsilon_k}$. The
integral over the magnitude of the internal momentum vector yields a factor
of $m^*z=(1+\frac{\partial\Sigma^>}{\partial\epsilon_k})^{-1}$. Thus,
altogether, the short wavelengths fluctuations multiply the self energy
(\ref{sigexp}), which we denote by $\Sigma^0$, by
$1+\frac{\partial\Sigma^>}{\partial\epsilon_k}$.
The
effective mass then becomes,
\begin{equation}
m^*=m\frac{1-(1+\frac{\partial\Sigma^>}{\partial\epsilon_k})
\frac{\partial\Sigma^0}{\partial\omega}}
{1+\frac{\partial\Sigma^>}{\partial\epsilon_k}}
\label{vertexmass}
\end{equation}
At sufficiently
low frequencies, however, $\frac{\partial\Sigma^0}{\partial\omega}\gg 1$,
and $m^*$ is independent of $\Sigma^>$, i.e., the
diverging part of the effective mass is
not affected by  the short wavelength gauge fluctuations.  We conclude, then,
that the diagram leading to the divergence of the effective mass through a
small energy--momentum exchange with the gauge field is not renormalized
by large energy--momentum exchange. This conclusion suggests that
our expression for the diverging part of the
effective mass (\ref{meff}) is {\it exact} in
the limit of $\omega\rightarrow 0$.

Does our method take into account all diagrams contributing to the leading
divergence in the self energy? Given a self energy diagram,
in our method we single out
the gauge field propagator with the smallest momentum, $q_{min}$
(denoted by the dashed line in Fig. (\ref{vertexfig})),
and assume that
all the momenta carried by the other gauge field propagators
(denoted by the dotted lines in Fig. (\ref{vertexfig}))  are
much larger. We thus effectively exclude the portions of phase space where
two or more  wave-vectors are close to $q_{min}$. This exclusion
should not affect the coefficient of the divergent portion of $\frac
{\partial\Sigma}{\partial\omega}$ if the self energy has contributions from
a large range of momentum exchange. This is indeed the case for a
Coulomb interaction, where the divergence is only logarithmic, and
the momentum exchange goes up to $2k_F$.
The separation of wave-vector scales
leading to Eq. (\ref{vertexmass}) is therefore
specific to the case of Coulomb interactions.

The separation of length
scales we use is in the spirit of a renormalization group analysis, which has
been explicitly introduced by Nayak and Wilczek, and justified in the
Coulomb case \cite{Wilczek1}\cite{Wilczek2}. Nayak and Wilczek have also
obtained a logarithmic singularity in the scaling of the frequency
$vs.$ $|k-k_F|$ for the fermion propagator, but they have not
discussed the value of the coefficient or consequences for Fermi liquid
theory \cite{Wilczek3}.

The situation with the Landau interaction function $f$ is
considerably more complicated than that for the effective mass. Due to
the identity (\ref{fltwi}), if $m^*$ diverges at the Fermi level we must have
$f_1=\frac{2\pi}{m}$. The other interaction parameters will be
renormalized by the short wavelength gauge fluctuations, which are not
treated correctly in our approximation. Thus, we are not able to make
precise predictions for their behavior.
In the limit where the bare
mass $m$ is small, so that the cyclotron frequency is large compared to
the scale of electron--electron interaction, we expect the
ground state and low--energy excitations to be confined to the
{\it electronic} lowest Landau level. Therefore,
one might expect the energy cost
of small deformations of the composite fermions' Fermi surface,
Eq. (\ref{landau}),  to be
determined only by electron--electron interactions, and to be independent
of the cyclotron energy, i.e., of $m$. In actuality, we
 see two exceptions to that
rule. The first is a galilean boost of the Fermi surface (e.g.,
$n({\bf k})\propto k_x$). In that case
the energy cost is determined by $f_1$, which is, indeed, $m$--dependent.
The second is $n({\bf k})=n_0$, a uniform  expansion of the Fermi
surface. In that case the energy cost is determined by $f_0$, which is
related to the composite fermions compressibility. As discussed in
section (2), the latter depends on the bare mass even when
$m\rightarrow 0$. Consequently, we find that in the limit of
$m\rightarrow 0$ the first two Landau parameters,
 $f_0$
and $f_1$, should have a value of the order of ${1\over m}$.
We believe, however, that  for $l\ge 2$,
 the value of $f_l$ is determined by the
energy scale of electron--electron interaction. We also believe that
$\ {\rm lim}_{l\rightarrow\infty}f_l\ne 0$, so that $f(\theta)$ should
include a delta function component, whose amplitude is determined
by electron--electron interactions. At this stage, however, we are not able to
substantiate this argument by a detailed calculation.

In view of our caveats regarding the perturbative calculation of
the Landau function, one may ask is there a  parameter whose smallness
could make our approximations for $\Sigma$ and $\Gamma$ close to their
exact values. This question is not easy to answer, since  our results do
not constitute  the first few terms in a power expansion in a small
parameter --- The use of an
RPA gauge field propagator involves a summation of all orders of
perturbation theory. A small parameter in the sense defined above might be a
parameter whose smallness makes the self energy (\ref{sigexp}) small.
An obvious candidate
is $\tilde\phi$ since ${\rm lim}_{{\tilde\phi}\rightarrow 0}
\Sigma=0$.
However, our results do not become exact when ${\tilde\phi}=0$.
At ${\tilde\phi}=0$ the magnetic field $B=2\pi{\tilde\phi}{n}
=0$, and the problem is that of interacting electrons in the absence of a
magnetic field. In the latter, which is of course a well studied problem,
 the electron--electron interaction gives rise
to a wave--vector dependent self energy, which renormalizes the mass,
in contrary to the vanishing self energy in our calculation for
${\tilde\phi}=0$. Consequently, our unperturbed problem could not be that
of ${\tilde\phi}=0$ and finite $\frac{e^2k_F}{E_F}$, and has to be that
of non--interacting electron at zero magnetic field, i.e.,
${\tilde\phi}=e=0$. Our perturbative analysis should therefore apply
to the limit where both
${\tilde\phi}\rightarrow 0$ and $\frac{e^2k_F}{E_F}
\rightarrow 0$. For the self energy (\ref{sigexp}) to be small, thus making
the perturbative treatment  sensible, the ratio between these
two parameters should satisfy
 $\frac{{\tilde\phi}E_F}{e^2k_F}\rightarrow 0$.
In view of that, the approximation we have obtained for the Landau
function can be better understood. At any finite non--integer $\tilde\phi$,
the problem we deal with is that of anyons in a magnetic field. Due to the
anyonic statistics, the ground state  of that problem is not confined to
the lowest Landau level even as $m\rightarrow 0$\cite{Read}.
Thus, the bare mass affects the
Landau function of the transformed fermions. It is only at the particular
values of integer $\tilde\phi$ that when $m\rightarrow 0$ the ground
state is purely composed of lowest Landau level wave functions. Our
perturbative treatment of the Landau function is not good enough to
capture that feature.

An alternative procedure which has been used to investigate a related problem
of particles interacting with a gauge field is to introduce $N$ species of
fermions, and to analyze the problem in the limit of large $N$
\cite{Kim}\cite{Aim}.
The RPA gauge field propagator in this model is identical to
our $D_{11}$, and the interaction with the gauge field induces a
current--current interaction between the fermions. The leading term
in the large $N$ limit of that model are similar to the ones included
in our self--consistent Green's function approximation. This large $N$
procedure has not been applied directly to the Chern--Simons problem
discussed here, however.

In conclusion, our perturbative calculation indicates that the effective mass
of the composite fermions diverges logarithmically, according to Eq.
(\ref{meff}),
as the Fermi surface is approached,
while the Landau function approaches
 a delta function in $\theta$.
We believe that the former is an exact statement,
while the latter is only  approximate.
In the next two sections we study possible
consequences of these singularities on observable quantities.

\section{Linear response functions and collective modes for the
composite fermions}
Having derived  approximate values for the effective mass and the Landau
function, we now turn to analyze the fermions' collective modes, and their
linear response to a driving force.
The collective modes we refer to here
are the solutions of Eq. (\ref{eomnu}) with $F=0$, i.e., they are fictitious
modes which would occur if the Hartree interaction were removed.
As explained in section (2), some of these modes are unaffected by the
inclusion of the Hartree term.
Generally, in the absence of a driving force
equation (\ref{eomnu})
has a continuum of solutions for which $\omega/q<v_F^*$,
and might have a discrete spectrum of solutions for which $\omega/q>v_F^*$.
For any particular Landau function, the number of modes in the
discrete part of the spectrum is smaller or equal to the number of
non--zero Landau parameters. Within the framework used in the
previous section we have found
 that the Landau function of the composite fermions
is $f(\theta-\theta')=\frac{(2\pi)^2}{m}\delta(\theta-\theta')$,
i.e.,  its Landau parameters are all non--zero and equal. The  effect of
this particular Landau function turns out to be a trivial one:
consider the slightly more general case of
 $f(\theta-\theta')=2\pi {\tilde f}
 \delta(\theta-\theta')$. Substituting this Landau
function in Eq. (\ref{eomnu}) we find
that its only effect is to shift
 the Fermi velocity from $\frac{k_F}{m^*}$ to
$k_F(\frac{1}{m^*}+\frac{\tilde f}{2\pi})$.
Galilean invariance, however, relates $f$ and $m^*$
and requires $\frac{1}{m^*}+\frac{\tilde f}{2\pi}=\frac{1}{m}$ \cite{Nozieres}.
Thus,  we find that the only
effect of any delta function $f(\theta)$
is to cancel the renormalization of the
Fermi velocity
and shift it back to its bare value. In the particular case we consider,
$m^*=\infty$ and $f(\theta)=\frac{(2\pi)^2}{m}\delta(\theta)$. The Landau
function shifts the mass back from infinity to its bare value.
The effective mass $m^*$ then cancels from  Eq. (\ref{eomnu}), and,
consequently,
 it does not affect the long wavelength low frequency
behavior of linear response function.
This result is of course consistent with the analyses of Kim {\it et.al.}
\cite{Kim}
and Altshuler {\it et.al.} \cite{Aim}
for the response functions.

The above results  could be viewed as a
limiting case of the behavior of the discrete spectrum when the number
of appreciably non--zero Landau parameters is very large:
the larger this number is,
the larger is the number of modes at the range
$\omega/q>v_F^*$, and the
fuzzier is the distinction between the continuum and the discrete
spectra. When all Landau parameters are appreciably non--zero, the
 number of modes at  $v_F^*<\omega/q<v_F$ becomes infinite,
and the discrete spectrum may become
continuous.

Our use of the equation of motion (\ref{eomnu}) for a case in which the
effective mass diverges at the Fermi level might be too crude an approximation.
One may expect a more careful analysis to introduce a frequency dependence to
$m^*$ and $f$ in that equation, as presented in Eq. (\ref{ftheta}).
However, at low frequencies, the main
feature we stressed here should not change. The Landau function has many
appreciably non--zero Landau parameters, and thus the number of discrete
collective modes is very large. At finite temperature or
finite wavelengths these modes acquire a
width, due to quasi--particle scattering, and the discrete spectrum becomes
more of a continuum.

We have seen that the singularities
of $m^*$ and $f$ mutually cancel in even denominator
$q,\omega\rightarrow 0$ linear response
functions at zero temperature.
This might cast some doubt on the physical significance of the divergence of
the effective mass. Are there any observable quantities in
which this divergence is  manifested?  The next section is devoted to the
study of one such manifestation, in
the  variation of the chemical potential with the density at and
between fractional quantum Hall states.

\section{The variation of the chemical potential with the density at and
between fractional quantum Hall states}

We now turn to consider the case of a large, yet finite, $p$, i.e.,
electrons at fractional quantized Hall states close to
$1/\tilde{\phi}$. The question to be discussed is
the variation of the transformed fermions
chemical potential $\mu$ as a function of
the density $n$. For non--interacting electrons in a constant
magnetic field, the chemical
potential jumps discontinuously when the density is varied.
The jumps take place at integer filling factors, and their height,
$\hbar\omega_c$,
 is the energy gap for the integer quantized Hall effect states.
For the problem we consider, care should be taken in defining an
analogous question, since the density can be varied with the external
magnetic field $B$
being kept fixed, or with the effective magnetic field $\Delta B$ being kept
fixed. The different significance of the two quantities
deserves further elaboration.
The chemical potential $\mu$
is the energy cost involved in adding a transformed fermion to the
system (along with a compensating contribution to the charge of the
uniform positive background), while $\Delta B$ is held fixed.
This chemical potential should be distinguished from the chemical
potential $\mu_e$ for electrons, defined as the change in the ground state
energy when one electron is added to the system and the external field $B$
is held fixed. Since the total number of transformed fermions equals the total
number  of
electrons $N_e$, we can simply write,
\begin{equation}
\begin{array}{ll}
\mu=\frac{\partial E}{\partial N_e}|_{\Delta B}\\ \\
\mu_e=\frac{\partial E}{\partial N_e}|_{B}
\end{array}
\label{mudef}
\end{equation}
where $E$ is the system's ground state energy.

Suppose that a fermion is added to the ground state at
$\nu=\frac{p}{\tilde{\phi}p+1}$, by use of the transformed fermion creation
operator $\Psi^+({\bf r})$, and suppose that the resulting wavefunction
is projected onto the lowest available energy state.
The added fermion carries  a flux tube of $\tilde\phi$ flux quanta,
in a direction
opposite to the external field $B$. If the latter is kept
constant, this flux tube is uncompensated and  the total  flux
becomes smaller by $\tilde\phi$ flux quanta.
Thus, $\tilde\phi$ fermions are pushed out of any one of
the $p$ occupied Landau  levels, and the $p+1$'th level receives a total of
$\tilde{\phi}p+1$ fermions.
If, however, it is $\Delta B$ that is kept constant, the
$\tilde\phi$ flux tubes carried by the fermion are compensated by
a change in $B$. Consequently,
the total flux does not change, and only one fermion
resides in the $p+1$ Landau level.  In that case, the resulting
 state has a fractional charge $-e^*$ in the vicinity of the point $\bf r$,
 where $e^*=\frac{e}{\tilde{\phi}p+1}$, while the remaining charge
 $-(e-e^*)$ is pushed to the boundary of the system; i.e., a quasi--particle
 is added to the interior of the system. It follows from this analysis
 that the energy gap $E_g(\nu)$, defined as the energy to add one
 quasi--particle and one quasi--hole (well separated from each other)
 to the ground state at $\nu=\frac{p}{\tilde{\phi}p+1}$, may be identified
 with a discontinuity in the fermion chemical potential $\mu$ at
 filling factor $\nu$.

Our approach in studying $\mu(n)$ for a fixed $\Delta B$
 is based on a perturbative calculation of the energy cost
involved in adding one fermion to the system in three different
situations: ({\it a}) when $p-1$ levels are completely filled, and the added
fermion is the first one in the $p$'th level; ({\it b}) when the $p$'th
level is almost filled, and the added fermion occupies the last vacant state
on that level; ({\it c}) when $p$ levels are completely filled and the
added fermion is the first one in the $p+1$'th level. The three are denoted
$\mu_a(p),\mu_b(p),\mu_c(p)$, respectively. Obviously, $\mu_a(p)=\mu_c(p-1)$.
In the absence of the perturbation, for any given $p$,
$\mu_a=\mu_b$, and
$\mu_c-\mu_b=\Delta\omega_c\equiv \frac{\Delta B}{m}$. In the presence of
the perturbation the  jump in the chemical potential, $\mu_c-\mu_b$, will be
denoted by $\Delta\omega_c^*$.

Our calculation of the three chemical potential differences is based on
a perturbative study of the fermion propagator, defined in Eq. (\ref{gfdef}).
One comment regarding the exact propagator
is in place before we turn to perturbation theory.
Since $\Delta B\ne 0$, the Hamiltonian is not invariant to translations,
and the fermion propagator is not diagonal in momentum representation.
As we show below, however,
{\it it is diagonal in Landau level representation}.

The
fermion propagator in Landau level representation is defined by
\begin{equation}
G(n',m',t;n,m,t')\equiv\int d^2r\int d^2r' \phi_{n',m'}^*({\bf r'})
{\tilde G}({\bf r},t,{\bf r'}, t') \phi_{n,m}({\bf r})
\label{llrep}
\end{equation}
where $\phi_{n,m}({\bf r})$ is the wavefunction for the state with
angular momentum $m$ and Landau level index $n$, and a symmetric gauge
is used. As we now turn to show,
$G(n',m';n,m)$ is non--zero  only for $m=m'$ and $n=n'$. The argument for
that is based on a spherical geometry \cite{Haldane}. As discussed by
Haldane, the problem of electrons in a magnetic field on a plane can be
studied by using a spherical geometry and taking the limit of an
infinite sphere. Similar to the plane, the single particle eigenstates
of electrons on a sphere, subject to
 a magnetic field perpendicular to the sphere,
are classified to energy--degenerate
Landau levels.
Unlike the plane, all states in the
$n$'th Landau level are eigenstates of the angular momentum operator
${\bf L}^2$, with the eigenvalue $(S+n)(S+n+1)$,
where $2S$ is the number of flux quanta threading the sphere.
 States within a
Landau level can be taken to be
 eigenstates of $L_z$, the $z$--component of the angular
momentum, with integer
eigenvalues ranging between $-S-n$ to $S+n$. The spherical analog
to a translationally invariant state is a state of zero
total angular momentum.
As usual, $G(n',m',t;n,m,t')$ is the amplitude of evolution from an initial
state to a final state. Here, one state is a state in which
 a particle (or a hole)   with
 angular momentum $S+n$ and a
$z$--component $m$ is added to  a system that had previously
no angular momentum. The other state is a state in which a particle
(or a hole) with
an angular momentum $S+n'$
and a $z$--component $m'$ is added to  a system that
had previously
no angular momentum. Both states are eigenstates of the angular
momentum operators ${\bf L}^2$ and $L_z$.
 Since the Hamiltonian conserves angular momentum, the amplitude for evolution
from one state to the other is zero unless $l=l'$ and $m=m'$.

We shall begin with a calculation restricted to a lowest--order--in--$D$
approximation, analogous to that employed in section (3b) for the study of
the effective mass $m^*$.
We shall see that
in the limit of large $p$ this calculation follows closely the
calculations of the self energy and the effective mass for $\Delta B=0$.
In particular, similar to the $\Delta B=0$ case,
the self energy at the large $p$ case is predominantly a function of frequency.
At the end of the section we shall discuss corrections to the
lowest--order--in--$D$ approximation, and argue that the asymptotic form
of the energy gap $\Delta\omega_c^*$ for large $p$ is unaffected by any
corrections.

The effect of the
perturbation is studied through
the self energy $\Sigma$. In the first--order--in--$D$ approximation,
$\Sigma$ is calculated by use of the free fermion propagator, given by,
\begin{equation}
G_f^{(0)}(\omega, l)=\frac{1}{\omega+\mu-(l-1/2)\Delta\omega_c+
i\eta{\rm sgn}(\omega)}  \ \ {\mbox {\rm for}}\ \  \Delta B\ne 0
\label{fermiongfb}
\end{equation}
where $\Delta\omega_c\equiv\Delta B/m$,  $m$ is the
bare mass,  $\mu$ is the chemical potential for the transformed fermions,
 and $l$ is an index for the Landau level of
the transformed fermion.

The self energy of a state in the $l$'th
Landau level is, in principle, a function of frequency, of the chemical
potential (i.e., of the number of filled levels) and of $l$. It is given,
in our first--order--in--$D$ approximation,  by
\begin{equation}
\Sigma(l,\omega,\mu)=\sum_{j=1}^\infty
i\int\frac{d^2q}{(2\pi)^2}\int \frac{d\Omega}{2\pi} {\cal M}(l,j;q)
\frac{1}{\Omega+\omega+\mu-\epsilon_j+i\eta{\rm sgn}(\Omega+\omega)}
D({\bf q},\Omega) \\ \\
\label{sigma}
\end{equation}
In the above equation, $\mu$ is a chemical potential such that
$p$ levels are filled, $\epsilon_j\equiv(j-1/2)\Delta\omega_c$ and
\begin{equation}
{\cal M}(l,j;{\bf q})
\equiv{1\over N_\phi}\sum_{g,g'}\big|\langle l,g|J_\perp
({\bf q})|j,g'\rangle\big|^2
\label{mljq}
\end{equation}
where $N_\phi$ is the Landau level degeneracy,
$j$ is a Landau level index, $g,g'$ are indices of states within the
Landau level and  $J_\perp$ is the transverse current operator.

In the following calculation of
 the self energy $\Sigma$ we approximate  the
gauge field propagator $D({\bf q},\omega)$
in (\ref{sigma}) by Eq. (\ref{gaugegf}), i.e.,
by its value at $p=\infty$.  We denote the latter by
$D^\infty$. We believe that this is a good
approximation since  for a large $p$ most of the contribution to the
$q$ integral comes from
$q>R_c^{-1}$, where $R_c$ is the cyclotron radius corresponding to the
motion of a composite fermion in the $p'th$ Landau level in a magnetic field
$\Delta B$. In that range of wave--vectors,
the non--interacting
 fermions' response function (the fermions' free bubble)
 is relatively insensitive to the
magnetic field. Since the dependence of the RPA gauge field propagator on
$p$ is only through its dependence on the fermions' response function,
we expect the gauge field RPA propagator to
be insensitive to $\Delta B$, too.

In terms of the self energy
(\ref{sigma}),
\begin{mathletters}
\label{mus}
\begin{equation}
\mu_c-\mu_b=\Delta\omega_c+\Sigma(p+1,\Delta\omega_c^*,\mu_{b}(p))-
                           \Sigma(p,0,\mu_{b}(p))
\label{musa}
\end{equation}

\begin{equation}
\mu_c-\mu_a=\Delta\omega_c+\Sigma(p+1,\Delta\omega_c^*,\mu_{b}(p))-
                                  \Sigma(p,\Delta\omega_c^*,\mu_{b}(p-1))
\label{musb}
\end{equation}

\begin{equation}
\mu_b-\mu_a=\Sigma(p,0,\mu_b(p))-\Sigma(p,\Delta\omega_c^*,\mu_b({p-1}))
\label{musc}
\end{equation}

\end{mathletters}

We start by calculating  the energy gap
$\Delta\omega_c^*=\mu_c-\mu_b$, and, in particular,
its dependence on $p$. The two self energies appearing in
Eq. (\ref{musa}) differ in their Landau  level index and
frequency arguments, but have the same number of filled Landau levels.
The self energy depends on the Landau level index only through the matrix
elements, and this dependence is weak. We can therefore approximate
\begin{equation}
\begin{array}{ll}
\Delta\omega_c^*\approx \Delta\omega_c(1+\frac{\partial\Sigma}{\partial
\epsilon_p})+
\sum_{j=1}^\infty i\int &\frac{d\Omega}{2\pi}
\int\frac{d^2q}{(2\pi)^2} {\cal M}(p,j;q)\Delta\omega_c^*
\\ \\
&
\times\frac{1}{\Omega+\Delta
\omega_c^*+\mu-\epsilon_j+i\eta{\rm sgn}(\Omega+\Delta\omega_c^*)}
\frac{1}{\Omega+\mu-\epsilon_j+i\eta{\rm sgn}(\Omega)}
D^\infty({\bf q},\Omega)
\end{array}
\label{diffmult}
\end{equation}

In the limit $p\rightarrow\infty$, $\frac{\partial\Sigma}{\partial\epsilon_p}$
approaches its value in the case of $\Delta B=0$. As we saw in
Section (3b), in our calculation of $m^*$ in the $\Delta B=0$ case,
 this term could be neglected within the lowest--order--in--$D$
 approximation, since the $\omega$--dependence of the self energy
 is more singular
when  $\omega\rightarrow 0$. We make the same approximation here for large $p$.

Since the gauge field propagator $D^\infty({\bf q},\Omega)$ depends
smoothly on $\Omega$ and the matrix
elements ${\cal M}(p,j;q)$ depend smoothly on $j$, we can replace the sum over
$j$ by an integral $\frac{1}{\Delta\omega_c}
\int d\epsilon_j$. Most of the contribution comes from $\epsilon_j\approx \mu$,
and the limits of integration can be extended to include all the real axis.
Carrying out this integration, we get,
\begin{equation}
\Delta\omega_c^*=\Delta\omega_c
+\frac{1}{\Delta\omega_c}
\int\frac{d^2q}{2\pi}\int_0^{\Delta\omega_c^*} \frac{d\Omega}{2\pi}
{\cal M}(p,{\tilde j};q)
D({\bf q},\Omega)
\label{almostgap}
\end{equation}
where ${\tilde j}=\frac{\mu+\Omega}{\Delta\omega_c}\approx p$.

Thus, the next step is to calculate ${\cal M}(p,p;q)$.
 Assuming that ${\bf q}||{\hat x}$, $J_\perp(q)= v_y e^{iqx}$. The operators
$v_y,x$ can be expressed in terms of inter Landau level creation and
destruction operators $a_d^+,a_d$ and intra Landau levels creation and
destruction operators $a_g^+, a_g$ as \cite{Cohen}
\begin{equation}
\begin{array}{ll}
x={1\over\sqrt{2}} l_\Delta(a_d^++a_d+a_g^++a_g)\\
v_y={1\over\sqrt{2}}\omega_cl_\Delta(a_d^++a_d)
\end{array}
\label{dagger}
\end{equation}
where $l_\Delta\equiv 1/\sqrt{\Delta B}$
 is the magnetic length corresponding to
a magnetic field $\Delta B$.
In terms of these operators,
\begin{equation}
{\cal M}(p,p;q)=\left|\omega_c \frac{\partial}{\partial q}\{\langle p|
e^{i\frac{ql_\Delta}{\sqrt{2}}(a_d^++a_d)}|p\rangle \}
\right |^2
\label{me}
\end{equation}
The sum over $g,g'$ cancels the $N_\phi$ in the denominator of
Eq. (\ref{mljq}).

Eq. (\ref{me}) contains a derivative with respect to  $q$ of an  expectation
value of an operator at an harmonic oscillator energy eigenstate.
Using  $R_c= l_\Delta\sqrt{p-1/2}$, we get, in the limit
$p\rightarrow\infty$,
\begin{equation}
\langle p|e^{iqx}|p\rangle=
\frac{{\Delta\omega_c}}{2\pi}\int_0^{\frac{{\Delta\omega_c}}{2\pi}}
dt e^{iqR_c\sin{{\Delta\omega_c} t}}=J_0(qR_c)
\end{equation}
where $J_0$ is Bessel's function.
Consequently,
\begin{equation}
\begin{array}{ll}
{\cal M}(p,p;q)=({\Delta\omega_c} R_c J_1(qR_c))^2\approx\left\{
\begin{array}{ll}
({\Delta\omega_c} R_c)^2(\frac{qR_c}{2})^2
\ \ \mbox{\rm for} \ \ qR_c\ll 1 \\ \\
({\Delta\omega_c} R_c)^2
\frac{2}{\pi qR_c}\cos^2({qR_c-0.75\pi})\ \ \mbox{\rm for}
\ \ 1\ll qR_c\ll R_c/l_\Delta \\ \\
{\rm  negligible\ \ for\ \ } ql_\Delta\gg 1
\end{array}\right.
\end{array}
\end{equation}
Substituting into (\ref{almostgap}), and observing that
in the large $p$ limit most of the contribution comes from the intermediate
wave-vector regime, we find
\begin{equation}
\Delta\omega_c^*\approx  \Delta\omega_c
+v_F\int_{R_c^{-1}}^{l_\Delta^{-1}}
 dq\int_0^{\Delta\omega_c^*} \frac{d\Omega}{2\pi}
\frac{1}{\pi }
D({\bf q},\Omega)
\end{equation}
where we used $\Delta\omega_c R_c=v_F$. Thus, the energy gap for the large $p$
limit is,
\begin{equation}
\Delta\omega_c^*(p)= \frac{e^2k_F\pi}{\epsilon
\tilde{\phi}(\tilde{\phi}p+1)}\frac{1}
{\ln{(2p+1)}}
\label{gap}
\end{equation}
This expression for the energy gap is equivalent to the one given in the
introduction, Eq. (\ref{gapp}) (note that for large $p$, $\ln{(2p+1)}
\sim\ln{(\tilde{\phi}p+1)}\sim \ln{p}$, and
$k_F\sim l_H^{-1}(2/\tilde{\phi})^{1/2}$). It is also consistent
with the recent calculation of Kim {\it et.al.}\cite{Kimstamp}.
The origin of the logarithmic singularity is the same as its origin in
the calculation of the effective mass for the $\nu=1/\tilde{\phi}$ case,
namely, the $1\over q$
divergence of $D^\infty(q,0)$. In fact, after making the approximation of
the sum over $j$ by an integral (Eq. (\ref{almostgap})), we brought the
self energy in the large $p$ case to a form very similar to that of
the self energy in the $\Delta B=0$ case, with the main difference being
the cut--off of the infra--red divergence in the momentum integral,
resulting from the suppression of the matrix elements for $q\ll R_c^{-1}$.
Thus,
{\it the magnitude of the discontinuous jump in the chemical potential at
the fractional quantized Hall states reflects the logarithmic
singularity of the fermions' effective mass.
For any finite $p$, this singularity is cut--off by
$\ln{\frac{R_c}{l_H}}\sim\ln (2p+1)$. }

Having calculated the energy gap, we now turn to $\mu_c(p)-\mu_a(p)$.
The
sum of matrix elements, ${\cal M}(l,j;q)$, depends strongly
on $|l-j|$, and only weakly on $l,j$ independently.  Consequently,
the self energy difference appearing in Eq. (\ref{musb})
is regular, and $\mu_c(p)-\mu_a(p)\approx \Delta\omega_c$. Similarly,
$\mu_b(p)-\mu_b(p-1)\approx\Delta\omega_c$.

Lastly, we turn to $\mu_b-\mu_a$, the last
 of the three chemical potential differences
defined in Eqs. (\ref{mus}). Obviously, it is
 uniquely determined by Eqs. (\ref{musa}) and (\ref{musb}).
It is, however, interesting to carry out an independent
calculation of this difference in chemical potentials.
Unlike the self energy difference
$\mu_c-\mu_b$,  the self energy difference appearing in
Eq. (\ref{musc}) involves self energies at different chemical
potentials, i.e., different number of filled levels. It is given by,
\begin{equation}
\begin{array}{lll}
\mu_b-\mu_a= \\   \\
i\int\frac{d^2q}{(2\pi)^2}\Big[\sum_{j=1}^{p}{\cal M}(p,j;q)\int
\frac{d\Omega}{2\pi}
\Big(\frac{\Delta\omega_c-\Delta\omega_c^*}
{\Omega+\mu_{b}(p)-\epsilon_j+i\eta}
\frac{1}{\Omega+\Delta\omega_c^*+\mu_b(p-1)
                                -\epsilon_j+i\eta}\Big)
D({\bf q},\Omega)
\\ \\
-\sum_{j=p+1}^\infty
{\cal M}(p,j;q) \int \frac{d\Omega}{2\pi}
\Big(
\frac{\Delta\omega_c-\Delta\omega_c^*}
                 {\Omega+\mu_b(p)-\epsilon_j-i\eta}
\frac{1}{\Omega+\Delta\omega_c^*+\mu_b(p-1)-\epsilon_j-i\eta}
\Big)
D({\bf q},-\Omega)\\ \\
+2
{\cal M}(p,p;q) \int \frac{d\Omega}{2\pi}
\frac{1}{\Omega+\Delta\omega_c^*+\mu_b(p-1)-\epsilon_p-i\eta}
D({\bf q},\Omega)\Big ]& \\ \\
\approx \frac{\partial\Sigma}{\partial\omega}(\mu_b-\mu_a)
+\int\frac{d^2q}{(2\pi)^2} {\cal M}(p,p;q)D({\bf q},\epsilon_p
-\Delta\omega_c^*-\mu_b(p-1))
\end{array}
\label{comp}
\end{equation}
In writing this expression we have separated the difference between the two
self energies to a part that does not involve a change in the level
occupation and a part that describes the change of the $p$'th level
from vacant to occupied. The first part varies smoothly when the
frequency and chemical potential are varied, and is therefore
approximated by a derivative. In the second part, in the large $p$ limit,
$D({\bf q},\epsilon_p
-\Delta\omega_c^*-\mu_b(p-1))$ can be replaced by
$D({\bf q},0)$. Following these approximations, we
indeed find  that
\begin{equation}
\begin{array}{ll}
\mu_b-\mu_a&=
\Delta\omega_c \frac{
\frac{\tilde{\phi}^2}{2\pi}
\frac{\epsilon k_F}{e^2 }\ln{(2p+1)}
}
{m+\frac{\tilde{\phi}^2}{2\pi}
\frac{\epsilon k_F}{e^2 }\ln{(2p+1)}}
\\ \\
&=
\Delta\omega_c-\Delta\omega_c^*
\label{slanting}
\end{array}
\end{equation}

Eqs. (\ref{gap}) and (\ref{slanting}) confirm the  picture
of $\mu(n)$ for fixed $\Delta B$
given in Fig. (\ref{mufig}): the chemical potential jumps discontinuously
whenever the fermions start occupying a previously vacant level. The
magnitude of that jump is singularly reduced by the interaction
with the gauge field fluctuations. In between two discontinuous jumps,
the chemical potential is not constant. The total change of the chemical
potential during the population of a level is
$\Delta\omega_c-\Delta\omega_c^*$, and it approaches $\Delta\omega_c$ as
$p\rightarrow\infty$, i.e., as the discontinuous jump is suppressed.
Our calculation does not give any information about the way the chemical
potential varies between $\mu_a$ and $\mu_b$. One could imagine a
continuous linear variation, which would result from a Hartree--Fock
approximation, or a discontinuous variation, due to, e.g., the formation of
fractional quantum Hall states of the composite fermions. Considering
the fact that the fractional quantum Hall effect has only been observed
in the first few Landau levels, and the fact that states in high Landau levels
are extended over a radius of $R_c\gg l_\Delta$,
we believe that a linear variation is at least an excellent
approximation.

Thus far, we have restricted our analysis to a lowest--order--in--$D$
approximation for the self energy at $\Delta B\ne 0$. We found that the
energy gap $\Delta\omega_c^*$ obtained in this approximation is
consistent with Eq. (\ref{gapintro}) and with the logarithmically
diverging mass obtained by a similar approximation for the $\Delta B=0$
case (see Section (3b)). In Section (3c) we argued that the calculated
logarithmic divergence of $m^*$, together with its coefficient, are
in fact exact in the $\omega\rightarrow 0$ limit. This is because a
Ward identity guarantees that corrections to the vertices due to
short wavelength fluctuations are cancelled by the correction to the
intermediate fermion Green's function (see Fig. (\ref{vertexfig}))
and by the factor $(1+\frac{\partial\Sigma}{\partial \epsilon_p})^{-1}$,
which enters the expression for $m^*$ (Eq. (\ref{effmass})).
A similar cancelation takes place when we extend our calculation of
the energy gap $\Delta\omega_c^*$ beyond the lowest--order--in--$D$
calculation. Again, we find that corrections to the vertices are
cancelled by corrections to the intermediate fermion Green's function and
by the factor $(1+\frac{\partial\Sigma}{\partial \epsilon_p})$ neglected
in Eq. (\ref{diffmult}). Thus, we find that Eq. (\ref{gap}) remains exact
in the limit $p\rightarrow\infty$. Due to the similarity with the analysis
presented in Section (3c), we omit the details here.

To conclude this section, we make one more comment regarding the energy gap.
Our calculation of the FQHE energy gaps is based on the study of poles of
a single fermion Green's function. An alternative way to extract the
energy gap
$\Delta\omega_c^*$
is through the linear response functions $\Pi$ and $K$ at
an integer fermion's filling factor $p$. As illustrated by an RPA
calculation \cite{Hlr} and a modified RPA calculation \cite{Simon},
in the {\it large wave-vector} limit,
 $q\gg 2k_F$, the lowest poles of both $\Pi$
and $K$ occur at a frequency that equals the energy gap. These poles
describe the excitation of a well separated quasi--hole quasi--particle
pair. The modified RPA calculation employs an approximation in which
only one Landau parameter, $f_1$, is non--zero, and shows that while
at the large wave-vector limit the excitation energies are unaffected
by the Landau parameter, this is not the case for small values of $q$.
In the latter case the quasi--particle and quasi--hole are not well
separated, and their mutual interaction affects the energy of the excitation.

\section{Fermion Green's function and coupling to $D_{00}$}

As has been noted above, the logarithmic divergence of $\frac{\partial\Sigma}
{\partial\omega}$, arising from interactions with the gauge field, implies
that the weight $z_k$ of the quasi--particle contribution to the fermion
Green's function $G_f$ is proportional to $1/\ln|k-k_F|$ when
$k\rightarrow k_F$, and vanishes at the Fermi surface. However, as
originally noted by HLR, there is another effect which causes $G_f$ to
vanish for any value of $\nu$, for any momentum and any finite energy, in
the limit of infinite system size. This effect arises from of the coupling
of the fermion propagator to the longitudinal gauge propagator $D_{00}$,
which thus far has been neglected in our discussion.

The physical reason for this divergence is that the transformed fermion
operator $\Psi^+({\bf r})$ has the effect of adding an electron at
point $\bf r$, and instantaneously turning on a solenoid containing two flux
quanta. The impulse electric field generated by the solenoid produces
inter Landau level excitations at large distances from $\bf r$ (i.e., it
produces long--wavelength magnetoplasma excitations at the cyclotron
frequency $\omega_c$) whose mean number diverges logarithmically with
the size of the system. The probability that the many electron state has
remained entirely in the lowest Landau level (or in any other state with
only a finite number of cyclotron excitations) is therefore found to
vanish as a power of the size of the system \cite{Hlr}.

In order to have a non vanishing weight for the one fermion Green's
function at low energies, it is necessary to define a renormalized version
of the Green function. One possibility would be to carry out calculations
for a large but finite system, and to renormalize the overall weight by
a size--dependent factor before taking the limit of infinite size.
An alternative approach is to work in an infinite system in the
limit where all operators are projected onto the lowest Landau level
{\it of the electrons}. This is equivalent to taking the limit where
the bare mass $m\rightarrow 0$ (i.e., $\omega_c\rightarrow\infty$),
while the electron--electron interaction is held constant.
Simultaneously, the flux in the solenoid associated with
$\Psi^+({\bf r})$ should be turned on at a rate slow compared with
$\omega_c$ but fast compared to the scale of the electron--electron
interaction.

Although neither of these procedures has been carried out in detail,
we expect that the renormalized fermion Green's function should
indeed have the low energy properties obtained for $G_f$ in the
previous sections, where the coupling to $D_{00}$ was neglected.
In particular, the renormalized Green function should have a singularity
at $\omega=\frac{k_F}{m^*(\omega)}(k-k_F)$, whose weight vanishes as
$1/\ln{|k-k_F|}$. There is no indication that the coupling to $D_{00}$
would change the singular behavior of the quasi--particle dispersion
for $k\rightarrow k_F$, or of the energy gaps at
$\nu=\frac{p}{\tilde{\phi}p+1}$.

\section{Summary}

In conclusion, in this work we discussed the effect of the diverging long
wavelength low frequency fluctuations of the Chern--Simons gauge field
on the properties of the electronic state at and near $\nu=1/\tilde{\phi}$.

In Sec. (2) we constructed an extension of Fermi liquid theory designed to
describe the effect of the Chern--Simons fluctuations on the low energy
excitations of the composite fermions. We emphasized the separation between
the Hartree and non--Hartree part of the interaction.

In Sec (3) we derived approximate expressions
for  the Fermi liquid parameters, namely the fermions'
effective mass and the Landau function.
We
showed that at $\nu=1/\tilde{\phi}$ both behave singularly. The
effective mass  diverges as the Fermi surface is approached,
and the Landau function has a part that approaches
 a delta function at $\theta=0$. We argued that our approximation
for the effective mass becomes exact in the limit $\omega\rightarrow 0$.

In section (4)
We  showed that  at $\Delta B=0$ the
singularities in $m^*$ and $f(\theta)$ are  cancelled in the
linear response functions in the limit of low frequencies and
long wavelengths.

In sec. (5)
we  investigated the variation of the
chemical potential of the transformed fermions
when the density is varied and $\Delta B$
is kept fixed. We showed that within our perturbative treatment, the
energy gap of the fractional Hall states depends singularly on the
filling factor, but the overall slope of $\mu(n)$ is not subject to
that singularity.

Finally, in Sec. (6) we briefly reviewed the effect of the fermions
coupling to longitudinal gauge field fluctuations on their Green's
function. We argued that this coupling does not affect the
results obtained in the previous sections.

Even within the limited context of clean systems, to which we have constrained
ourselves in this paper, we left several important questions opened. We would
like to point out some of these. First, while we believe our
approximation for the effective mass of the fermions becomes exact at a certain
limit, we do not have a similar understanding of our approximation for
the Landau function. We consider it likely that our approximation for
the Landau function yields the correct type of singularity (a delta function),
but with an incorrect coefficient.

Second, the analysis of this paper has been explicitly confined to the case of
Coulomb interactions between the electrons (i.e., an interaction that
behaves as  $1/r$ at large separations. Extensions to a shorter range
interaction would be of considerable theoretical interest, but must be
carried out with care. Perhaps most importantly, care should be
exercised in  discussing the very notions of an
effective mass and Fermi liquid parameters, since the decay rate of the
quasiparticle might be comparable to its energy (in contrast to the
Coulomb interaction, where the real part of the energy is larger than
the imaginary part by a factor of $\ln\omega\ $, a factor that diverges
as the Fermi surface is approached).

Finally, it should also be noted that in most practical applications to
present experiments on electron systems in partially filled Landau level,
the logarithmic contribution to the FQHE energy gap is smaller than
the finite, non--singular, contributions arising from short wavelength
fluctuations in the gauge field, which we were not able to calculate
reliably. These contributions depend on the behavior of the
electron--electron interaction at short distances, and are therefore
"non--universal".

\vskip 1in
\acknowledgments
We are grateful to Y.B. Kim, P.A. Lee, X.G. Wen, P.C.E. Stamp
and A. Furusaki for discussing their results with us prior to
publication, and we
are indebted to A.J. Millis, S. Simon and D. Chklovskii for
instructive discussions. This work was supported in part by NSF grant DMR
 94--16910.
AS acknowledges financial support of the Harvard Society of Fellows.

\appendix
\section{
Consistency in the approximations for the effective mass and the
Landau function}
In this appendix we show that the approximation scheme we have used
for approximating the Landau function does satisfy the first of the Fermi
liquid identities given in (\ref{fltwi}). Consistency with the second Ward
identity can be shown using similar methods. The first identity is
\begin{equation}
\frac{d\mu}{dn}=\frac{2\pi}{m^*}+f_0
\label{goal}
\end{equation}

As described in section (3),
we write the zeroth moment of the Landau function as
 \cite{Pita}\cite{Nozieres},
\begin{equation}
f_0={\rm lim}_{\omega\rightarrow 0}
{{-2\pi i}\over {
m^*(\omega)}}\int \frac{d^2k' }{(2\pi)^2} \int \frac{d\Omega}{2\pi}
{\tilde\Gamma}^{\epsilon}({\bf k},\omega;{\bf k'}\Omega;{\bf k'},
\Omega; {\bf k},\omega)
\left[\{G^2({\bf k},\Omega)\}_\epsilon-
\{G^2({\bf k},\Omega)\}_{\bf q}\right ]
\label{fefes}
\end{equation}
where
\begin{equation}
\begin{array}{ll}
\left\{G^2({\bf k},\omega)\right\}_\epsilon\equiv
{\rm lim}_{\epsilon\rightarrow 0}
G_f({\bf k},\omega+\epsilon)G_f({\bf k},\omega) \\ \\
\left\{G^2({\bf k},\omega)\right\}_{\bf q}
\equiv{\rm lim}_{{\bf q}\rightarrow 0}
G_f({\bf k+q},\omega)G_f({\bf k},\omega)
\end{array}
\label{refsapp}
\end{equation}
We would like to substitute approximate values for
${\tilde\Gamma}^\epsilon,G_f,
m^*$ in Eq. (\ref{fefes}), in order to obtain an approximation for
$f_0$. The choice of this approximate values, however, should be such
that the identity (\ref{goal}) is satisfied. We now turn to explain
how this choice should be made.

Our search for an approximation consistent with (\ref{goal}) starts with
expressing $f_0$ in terms of derivatives of the self energy $\Sigma$.
In doing that, we use a set of Ward
identities, one of which was already given above (\ref{ward}). We refer the
reader to Ref. \cite{Pita} for a comprehensive discussion of this set of
identities.
A shortened notation is useful here. We write Eq. (\ref{fefes}) as
$f_0=\frac{-2\pi i}{m^*}
\int[\{G^2\}_\epsilon-\{G^2\}_q]{\tilde\Gamma}^\epsilon$. Ward's identity
(\ref{ward}), which is useful here,
 is $\frac{\partial\Sigma}{\partial\omega}=
i\int {\tilde\Gamma}^\epsilon \{G^2\}_\epsilon$.
Also useful are the identities\cite{AGD}\cite{Pita}
\begin{equation}
{\tilde\Gamma}^\epsilon=
{\tilde\Gamma}^q-i\int {\tilde\Gamma}^\epsilon[\{G^2\}_\epsilon-\{G^2\}_q]
{\tilde\Gamma}^q
\label{gammas}
\end{equation}
(with $
{\tilde\Gamma}^q({\bf k},\omega;{\bf k'}\Omega;{\bf k'},
\Omega; {\bf k},\omega) \equiv
{\rm lim}_{q\rightarrow 0}{\rm lim}_{\epsilon\rightarrow 0}
{\tilde\Gamma}({\bf k},\omega;{\bf k'}\Omega;{\bf k'+q},
\Omega+\epsilon; {\bf k-q},\omega-\epsilon)$),
and the Ward identity for the chemical potential,
\begin{equation}
\frac{\partial\Sigma}{\partial\mu}=i\int {\tilde\Gamma}^q\{G^2\}_q
\label{wardmu}
\end{equation}

Substituting these three identities into Eq. (\ref{fefes}) we get,
\begin{equation}
\begin{array}{ll}
f_0&=\frac{2\pi}{m^*}
\frac{\partial\Sigma}{\partial\mu}-
\frac{2\pi}{m^*}\frac{\partial\Sigma}{\partial\omega}
+\frac{2\pi i}{m^*}
\int\int {\tilde\Gamma}^\epsilon
[\{G^2\}_q-\{G^2\}_\epsilon]{\tilde\Gamma}^q \{G^2\}_q
\\ \\
&=-\frac{2\pi}{m^*}
\frac{\partial\Sigma}{\partial\omega}+
\frac{2\pi}{m^*}\frac{\partial\Sigma}{\partial\mu}
+\frac{\partial\Sigma}{\partial\mu}f_0
\end{array}
\label{wards}
\end{equation}
leading to,
\begin{equation}
f_0=\frac{2\pi}{m^*}\frac{    \frac{\partial\Sigma}{\partial\mu}-
                              \frac{\partial\Sigma}{\partial\omega}}
                         {1-\frac{\partial\Sigma}{\partial\mu}}
\label{fzero}
\end{equation}
Note that all derivatives should be taken at the Fermi level.

Having expressed $f_0$ in terms of derivatives of the self energy,
we now express $\frac{d\mu}{dn}$ in similar terms.
The self energy is, generally, a function of ${\bf k}$ and $\omega$
(we deal with the $\Delta B=0$ case), and it depends also on the
chemical potential $\mu$. To stress that, we write it as
$\Sigma({\bf k},\omega;\mu)$. The chemical potential is the energy
needed for adding one fermion at the Fermi level. It is given by,
\begin{equation}
\mu=\epsilon_{k_F}+\Sigma(k_F,0;\mu)
\label{musigma}
\end{equation}
Taking the complete derivative $\frac{d\ }{dn}$ of both sides of
(\ref{musigma}), we get,
\begin{equation}
\begin{array}{ll}
\frac{d\mu}{dn}&=\frac{2\pi}{m}
            \frac{1+\frac{\partial\Sigma}{\partial\epsilon_k}|_{k=k_F}}
                 {1-\frac{\partial\Sigma}{\partial\mu}}  \\ \\
&=\frac{2\pi}{m^*}
            \frac{1-\frac{\partial\Sigma}{\partial\omega}}
                 {1-\frac{\partial\Sigma}{\partial\mu}}
\end{array}
\label{dmudn}
\end{equation}

Comparing with Eq. (\ref{fzero}) we see that
the approximate value for the effective mass used in Eq. (\ref{fefes})
to determine the Landau function should be the same as the one used in the
identity (\ref{goal}). Moreover, we get to the following prescription
for establishing an approximation for $m^*$ and $f$ that respects the
Fermi liquid identities (\ref{fltwi}):
\begin{itemize}
\item{
 The first step is to choose an approximation for the Green's function
 $G_f$, the interaction operator $\Gamma$ and the self energy $\Sigma$ in
 such a way that the Ward identities (\ref{ward}) and (\ref{wardmu}) are
 satisfied.}
\item{
 At the second step, the approximate value for $\Sigma$ should be used to
 extract an approximation for the effective mass.}
\item{
 Finally, the approximate values for the effective mass, the interaction
 operator and the Green's function should be substituted in Eq.
 (\ref{fgammab}), to obtain an approximate value for the Landau function.}

\end{itemize}

The application of this prescription
to the first--order--in--$D$ approximation we have used in Section (3) is
straightforward. As we discussed in Section (3) (below Eq. (\ref{ward}))
the Ward identity (\ref{ward}) is satisfied if the fermion Green function is
approximated by the free one $G_f^0$, the interaction operator by Eq.
(\ref{zerogamma})
and the self energy by (\ref{sezero}). Thus, in consistently approximating
the Landau function using Eq. (\ref{fefes}), these are the values that should
be used. In particular, the effective mass used should be the logarithmically
diverging one (\ref{meff}). Following these principles,
the approximation obtained for the Landau function is
the zero frequency limit of Eq. (\ref{ftheta}). The application of the above
prescription to the self--consistent Green's function approximation is
slightly more complicated, and is discussed in Appendix B.

To conclude, we note that the observations made above regarding the consistency
of approximations for the Landau functions are hard to make using the
conventional expression for the Landau function, Eq. (\ref{fgamma}). The
alternative equivalent expression we suggested, Eq. (\ref{fgammab}), makes
this analysis simpler.

\section{ Approximation of the interaction operator as a
sum of ladder diagrams.}
In this appendix we show that the self consistent Green's function
approximation for the self energy leads to the same
Landau function, Eq.  (\ref{ftheta}), as the one obtained from the
first--order--in--$D$ approximation. In the self consistent approximation, the
Green's function is $G\approx{[\Omega-\epsilon(k')-
\Sigma(k',\Omega)]^{-1}}=z(\Omega)(\Omega-v_F^*(k-k_F))^{-1}$,
where $z(\Omega)$
is the residue. As we saw in section (3), the leading singularity of the
self energy,
given by (\ref{sesc}), is the same as the one obtained in the
first--order--in--$D$ approximation.
The interaction operator should be approximated in
such a way that the Ward identity (\ref{ward}) is satisfied.
We start by finding out what this way is.

\begin{figure}[htbp]
\begin{center}
\leavevmode
\epsfbox{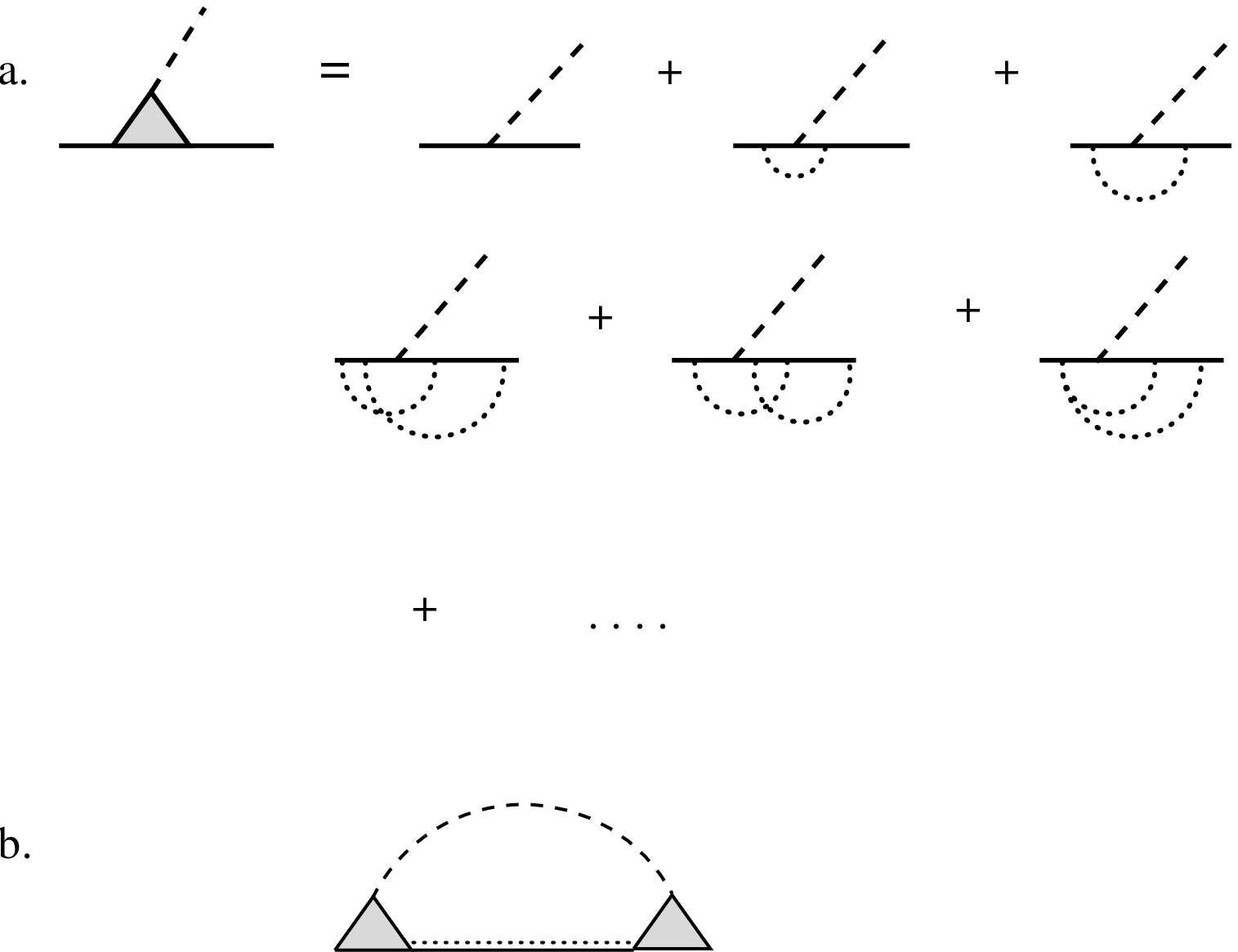}
\end{center}
\vskip 0.8truein
\caption{}{(a.) Some low order contributions to the
 dressed vertex. The dashed line represents a low frequency long wavelength
gauge field propagator. The dotted lines are short wavelength gauge field
propagators.
(b.) The renormalization of the self energy by short wavelength gauge
field fluctuations. The dashed line carries the lowest momentum in the
diagram. The double line (solid and dotted) denotes a fermion's Green's
function, renormalized by the short wavelength gauge field fluctuations.
As explained in text, the dominant contribution from the gauge field
propagator is not expected to be renormalized. }
\label{vertexfig}
\end{figure}

Taking the
derivative of (\ref{sesc}) with respect to frequency we get,
\begin{equation}
\begin{array}{ll}
\frac{\partial\Sigma}{\partial\omega}=
\int\frac{d^2k'}{(2\pi)^2}\int\frac{ d\Omega}{2\pi}
&\frac{1}{m^2}\frac{({\bf k}\times{\bf k'})^2}
       {|{\bf k}-{\bf k'}|^2}
\\ \\
&\times D_{11}(|{\bf k}-{\bf k'}|,
\Omega)
 \left (\frac{1}{\omega-\Omega-\epsilon(k')-
\Sigma(k',\omega-\Omega)}\right)^2
(1-\frac{\partial\Sigma}{\partial\omega}|_{\omega
-\Omega})
\end{array}
\label{recur}
\end{equation}
Substituting recursively $\frac{\partial\Sigma}{\partial\omega}$ into the
right hand side of (\ref{recur}) and comparing with Eq. (\ref{ward}), we
find that for Ward's identity to be satisfied, the interaction operator
should be approximated by the sum of ladder diagrams, illustrated in
Fig (\ref{vertexfig}).

Ward's identity requires $\frac{\partial\Sigma}{\partial\omega}=
i\int {\tilde\Gamma}^\epsilon \{G^2\}_\epsilon$. Since the left hand
side of the identity
is unchanged when the first--order--in--$D$ approximation is
changed to a self consistent one, so should be the case with the
right hand side.  In the self consistent
approximation the two Green functions carry two powers of the residue. The
integral over the magnitude of the intermediate momentum
 yields a factor of $\frac{1}{v_F^*}\sim m^*\sim z^{-1}$. Therefore,
for the right hand side of the Ward identity to be independent of $z$,
the interaction
operator should be $O(z^{-1})$. We now show that this observation is
consistent with the equation defining the ladder sum for the interaction
operator, and then turn to examine its implications on the approximation
for the Landau function.

The ladder diagram approximation for the interaction operator is the solution
to the equation,
\begin{equation}
\begin{array}{ll}
{\tilde\Gamma}^\epsilon&({\bf k},\omega;{\bf k'}\Omega;{\bf k'},
\Omega; {\bf k},\omega)=\frac{1}{m^2} \frac{({\bf k}\times{\bf k'})^2}
                                {|{\bf k}-{\bf k'}|^2}
                                D(|{\bf k}-{\bf k'}|,\omega-\Omega)\\ &
+\int \frac{d^2k''}{(2\pi)^2}\int\frac{d\omega''}{2\pi}
\frac{1}{m^2} \frac{({\bf k}\times{\bf k''})^2}
                                {|{\bf k}-{\bf k''}|^2}
                                D(|{\bf k}-{\bf k''}|,\omega-\omega'')
G^2(k'',\omega'')
{\tilde\Gamma}^\epsilon({\bf k''},\omega'';{\bf k'}\omega';{\bf k'},
\omega'; {\bf k''},\omega'')
\end{array}
\label{lad}
\end{equation}
If the interaction operator is to be $O(z^{-1})$, the first term
on the right hand side is negligible relative to the second. Neglecting that
term, equation (\ref{lad}) becomes an eigenvalue equation for ${\tilde\Gamma}$.
For this equation to be consistent with our $z$--power counting, the
second term of the right hand side should be of the same order of $z$ as
${\tilde\Gamma}$. This is indeed the case:  again, the two Green's
functions carry two powers of the residue, and
the integral over $k''$ introduces a factor of $(v_F^*)^{-1}
\sim m^*\sim z^{-1}$. Now,
our interest is in momenta $k,k'$
close to $k_F$ and frequencies $\omega,\Omega$ close to zero. The
denominators in $D$ and $G^2$ constrain $k''$ to be close to $k_F$ and
$\omega''\approx 0$. We may therefore approximate
$|k-k''|=2k_F\sin{\theta^{''}\over 2}$, where $\theta''$ is the angle
between $\bf k$ and $\bf k''$. The integral over $\theta''$ then diverges
at small angles, due to the divergence of $D$. This logarithmic
divergence is the same as the one leading to the divergence of $m^*$
and $z^{-1}$.    Altogether, then, the right hand side of (\ref{lad})
has the same order of $z$ as the left one, namely, the order
of ${\tilde\Gamma}$. We therefore conclude that $\tilde\Gamma\sim O(z^{-1})$.

Finally, we use the same type of
$z$--power counting to examine the approximation
for the Landau function. We start with Eq. (\ref{fgammab}).
Within the self--consistent Green's function approximation, the Green's
functions substituted in this equation carry two powers of the residue
$z$, and the interaction operator, as shown above, is of order $z^{-1}$.
Again, the integral over $k'$ introduces a factor of
$\frac{1}{v_F^*}\sim z^{-1}$, and the Landau function becomes $O(z^0)$,
i.e., it is the same as the one obtained in the first--order--in--$D$
approximation,  the zero frequency limit of Eq. (\ref{ftheta}).

\begin{references}
\bibitem[*]{byline} Future address (starting from October 1995): Department
of Condensed Matter Physics, Weizmann Institute of Sciences, Rehovot 76100,
Israel.
\bibitem{Hlr}
B.I. Halperin, P.A. Lee and N. Read, Phys. Rev. B {\bf 47}, 7312 (1993).

\bibitem{Fradkin}
A. Lopez and E. Fradkin Phys. Rev. B {\bf 44}, 5246 (1991).

\bibitem{Kz}
V. Kalmeyer and S.C. Zhang, Phys. Rev. B {\bf 46}, 9889 (1992).

\bibitem{Marston}
H.J. Kwon, A. Houghton and J.B. Marston, Phys. Rev. Lett. {\bf 73}, 284 (1994).

\bibitem{Kim}
Y.B. Kim, X.G. Wen, P.A. Lee and A. Furusaki, Phys. Rev. B {\bf 50} ... (1994).

\bibitem{Aim}
B.L. Altshuler, L. Ioffe and A.J. Millis, preprint.
\bibitem{Jain}
J. Jain, Phys. Rev. Lett. {\bf 63}, 199 (1989).
\bibitem{Wilczek1}
C. Nayak and F. Wilczek, Nuc. Phys. {\bf B417}, 359 (1994).
\bibitem{Wilczek2}
C. Nayak and F. Wilczek, preprint.
\bibitem{Gan}
J. Gan and E. Wong, Phys. Rev. Lett. {\bf 71}, 4226 (1993).
\bibitem{Misc}
J. Polchinski, Nucl. Phys. {\bf B422}, 617 (1994);
T. Holstein, R.E. Norton, and P. Pincus, Phys. Rev. B {\bf 8}, 2649 (1973);
M. Yu. Reizer, Phys. Rev. B {\bf 39}, 1602 (1989); {\it ibid.} {\bf 40},
11571 (1989); B. Blok and H. Monien, Phys. Rev. B {\bf 47}, 3454 (1993);
D.V. Kveshchenko, R. Hlublina and T.M. Rice Phys. Rev. B {\bf 48}, 10766
(1993); D.V. Kveshchenko and P.C.E. Stamp Phys. Rev. Lett. {\bf 71}, 2118
(1993), Phys. Rev. B {\bf 49}, 5227 (1994); P. Bares and X.G. Wen, Phys. Rev.
B {\bf 48}, 8636 (1993); Y.B. Kim and X.G. Wen, Phys. Rev. B {\bf 50},
8078 (1994); N. Nagaosa and P.A. Lee, Phys. Rev. Lett. {\bf 64}, 2550 (1990);
P.A. Lee and N. Nagaosa Phys. Rev. B {\bf 46}, 5621 (1992); L.B. Ioffe
and P.B. Wiegmann, Phys. Rev. Lett. {\bf 65}, 653 (1990); L.B. Ioffe and
G. Kotliar, Phys. Rev. B {\bf 42}, 10348 (1990)C.M. Varma {\it et.al.},
Phys. Rev. Lett. {\bf 63}, 1996 (1989); P.W. Anderson, Phys. Rev. Lett.
{\bf 65} 2306 (1990); {\it ibid.} {\bf 66}, 3226 (1991);
{\it ibid.} {\bf 67}, 2092 (1991); C.M. Varma, preprint; L. B. Ioffe, D.
Lidsky and B.L. Altshuler, Phys. Rev. Lett. {\bf 73}, 472 (1993);
L.B. Ioffe and A.I. Larkin, Phys. Rev. B {\bf 39}, 8988 (1989);
P.A. Lee, Phys. Rev. Lett. {\bf 63}, 680 (1989);
S. Chakravarty, R.E. Norton and O.F. Syljuasen Phys. Rev. Lett.
{\bf 74}, 1423 (1995).
\bibitem{Wilczek3}
An equation in Ref. (\cite{Wilczek1}), suggesting a scaling of the form
$\omega(\ln{\omega})^{1/2}\propto |k-k_F|$ has been corrected to the
relation $\omega\ln\omega\propto |k-k_F|$ in Ref. (\cite{Wilczek2}).
\bibitem{Nummist}
Unfortunately, there was a numerical error in the corresponding expression
published in Ref. \cite{Hlr}. For $\tilde{\phi}=2$ the coefficient in
Eq. (\ref{gapp}) differs by a factor of $1.23$ from the value quoted in
Ref. \cite{Hlr}.
\bibitem{Mahan}
G.D. Mahan, Many Particle Physics, Plenum Press, New York (1981), Sec. 3.3.

\bibitem{Pita}
E.M. Lifshitz and L.P. Pitaevskii, Statistical Physics, part 2, Pergamon Press.
See in particular sections 18-19 and 64-65.

\bibitem{AGD}
A.A. Abrikosov, L.P. Gorkov and I.E.
Dzyaloshinski, Methods of quantum field theory in
statistical physics, Dover publication, New York (1975)
\bibitem{Nozieres}
P. Nozieres, Theory of interacting Fermi systems, W.A. Benjamin,
New York, 1964. In particular, see chapters 1,5,6.
\bibitem{Read} N. Read, Semicon. Sci. Tech. {\bf 9}, 1859 (1994).
\bibitem{Haldane}
F.D.M. Haldane, Phys. Rev. Lett. {\bf 51}, 605 (1983).
See also S. Simon and B.I. Halperin, Phys. Rev. B {\bf 50}, 1807 (1994).
\bibitem{Cohen}
 C. Cohen--Tanoudji, B. Diu, F. Laloe,  Quantum Mechanics, Wiley, New York
(1977).

\bibitem{Kimstamp}
Y.B. Kim, X.G. Wen, P.A. Lee and P.C.E. Stamp, preprint.

\bibitem{Simon}
S. Simon and B.I. Halperin, Phys. Rev. B {\bf 48}, 17368 (1993).



\end{references}
\end{document}